\documentclass[pra,aps,%
 reprint,%
 twocolumn,
superscriptaddress,
nofootinbib,
longbibliography]{revtex4-2}
\usepackage{makecell} 
\usepackage{braket}
\usepackage{color}
\usepackage{mathtools}
\usepackage{dsfont}
\providecommand{\mathbbm}[1]{\mathds{#1}}
\usepackage[dvipsnames,table]{xcolor}
\usepackage{graphicx}
\usepackage{amsfonts}
\usepackage{amssymb}
\usepackage{booktabs}
\usepackage{amsmath}
\usepackage{amsthm}

\usepackage{textcomp}

\usepackage{array}
\usepackage{longtable}
\usepackage{enumitem}
\usepackage{verbatim}

\usepackage[T1]{fontenc}
\usepackage{etoolbox}

\usepackage{newtxtext}
\usepackage{newtxmath}
\usepackage[normalem]{ulem}

\makeatletter
\g@addto@macro\bfseries{\boldmath}
\makeatother

\usepackage[colorlinks=true,linkcolor=teal,citecolor=teal,urlcolor=teal,plainpages=false,pdfpagelabels]{hyperref}

\theoremstyle{definition}

\renewcommand{\qedsymbol}{$\blacksquare$}

\renewcommand{\qedsymbol}{\unskip\nobreak\quad\qedsymbol}
\renewcommand{\qedsymbol}{$\blacksquare$}

\DeclareMathOperator*{\argmax}{arg\,max}

\definecolor{hollywoodcerise}{rgb}{0.96, 0.0, 0.63}

\definecolor{dblue}{rgb}{0.0, 0.53, 0.74}

\definecolor{dgreen}{HTML}{26cc00}

\begin{document}

\title{Reinforcement learning of quantum circuit architectures\\for molecular potential energy curves}

\author{Maureen Krumtünger}
\thanks{MK and AW have contributed equally.}
\affiliation{Dahlem Center for Complex Quantum Systems, Freie Universität Berlin, 14195 Berlin, Germany}
\affiliation{Formerly at Porsche Digital GmbH, 71636 Ludwigsburg, Germany}

\author{Alissa Wilms}
\thanks{MK and AW have contributed equally.}
\affiliation{Dahlem Center for Complex Quantum Systems, Freie Universität Berlin, 14195 Berlin, Germany}
\affiliation{Porsche Digital GmbH, 71636 Ludwigsburg, Germany}

\author{Paul K. Faehrmann}
\affiliation{Dahlem Center for Complex Quantum Systems, Freie Universität Berlin, 14195 Berlin, Germany}

\author{\\Jens~Eisert}
\affiliation{Dahlem Center for Complex Quantum Systems, Freie Universität Berlin, 14195 Berlin, Germany}
\affiliation{Fraunhofer Heinrich Hertz Institute, 10587 Berlin, Germany}
\affiliation{Helmholtz-Zentrum Berlin für Materialien und Energie, 14109 Berlin, Germany}

\author{Jakob Kottmann}
\affiliation{{Institute for Computer Science, University of Augsburg, 86159 Augsburg, Germany }}
\affiliation{{Center for Advanced Analytics and Predictive Sciences, University of Augsburg, 86159 Augsburg, Germany }}

\author{Paolo Andrea Erdman}
\affiliation{Department of Mathematics and Computer Science, Freie Universität Berlin, 14195 Berlin, Germany}

\author{Sumeet Khatri}
\affiliation{Dahlem Center for Complex Quantum Systems, Freie Universität Berlin, 14195 Berlin, Germany}
\affiliation{Department of Computer Science and Center for Quantum Information Science and Engineering, Virginia Tech, Blacksburg, VA 24061, USA}

\date{\today}

\begin{abstract}

Quantum chemistry and optimization are two of the most prominent applications of quantum computers. Variational quantum algorithms have been proposed for solving problems in these domains. However, the design of the quantum circuit ansatz 
remains a challenge. Of particular interest is developing a method to generate circuits for any given instance of a problem, not merely a circuit tailored to a specific instance of the problem. 
To this end, we present a reinforcement learning (RL) approach to learning a problem-dependent quantum circuit mapping, which outputs a circuit for the ground state of a Hamiltonian from a given family of parameterized Hamiltonians. For quantum chemistry, our RL framework takes as input a molecule and a discrete set of bond distances, and it outputs a bond-distance-dependent quantum circuit for arbitrary bond distances along the potential energy curve. The inherently non-greedy approach of our RL method contrasts with existing greedy approaches to adaptive, problem-tailored circuit constructions. We demonstrate its effectiveness for the four-qubit and six-qubit lithium hydride molecules, as well as an eight-qubit H$_4$ chain. Our learned circuits are interpretable in a physically meaningful manner, thus paving the way for applying RL to the development of 
novel quantum circuits for the ground states of large-scale molecular systems. 

\end{abstract}

\maketitle

\section{Introduction}

One of the central challenges in quantum chemistry, both from a conceptual and application-oriented perspective, is the accurate computation of a molecule’s \emph{potential energy surface}, which describes its ground-state energy as a function of nuclear geometry. As a practically relevant simplification, the \emph{potential energy curve} (PEC) captures the ground-state energy along a single reaction coordinate, typically a bond distance, and serves as a valuable tool for analyzing reaction pathways, assessing molecular stability, and performing the simulation of molecular dynamics.

Obtaining chemically accurate PECs is challenging: while full configuration interaction (FCI) provides the exact ground state, its computational requirements scale exponentially with system size; consequently, more scalable approximations have been developed, typically at some cost in accuracy~\cite{jensen2007introduction}.

More recently, \emph{machine learning}  (ML) has become an indispensable tool for scientific development in the quantitative sciences, 
particularly in the prediction of molecular 
\cite{Molecular,Molecular2} and material characteristics \cite{Roadmap}, culminating in the 2024 Nobel Prize in Chemistry for AlphaFold, an \emph{artificial intelligence} (AI)-based tool for predicting protein structures~\cite{jumper2021AlphaFold}.  ML methods have also been increasingly employed to support PES and PEC calculations by learning the relationship between molecular structure and energy from training data \cite{PhysRevLett.98.146401,2025CoPhC.30809446T}. Alternative ML methods aim to directly predict the full electronic wave function~\cite {schutt2019quantumchemML}.

With the advent of quantum computers, new possibilities for computing PECs have emerged, placing quantum chemistry, particularly ground state energy estimation, among the first applications expected to benefit from their computational power~\cite{feynman1982simulation,hoefler2023disentangling,beverland2022assessing,scholten2024assessing,AIQT}.  
One of the most promising algorithms for estimating the ground state energy of a given Hamiltonian is the \emph{variational quantum eigensolver} (VQE) \cite{cerezo2021VQAreview, bharti2021noisy,Peruzzo_2014}.
In the VQE, a parameterized state vector, $\ket{\psi(\boldsymbol{\theta})}$, is prepared on a quantum computer, while its parameters $\boldsymbol{\theta}$ are iteratively optimized by a classical subroutine to 
minimize the energy:
\begin{align}
E_{\text{VQE}}(\hat{H})=\underset{\boldsymbol{\theta}}{\text{min}}\braket{\psi_0|\hat{U}^\dagger(\boldsymbol{\theta})\hat{H}\hat{U}(\boldsymbol{\theta})|\psi_0},\label{cost_function_vqe}
\end{align}
where $\hat{H}$ is the Hamiltonian and $\hat{U}(\boldsymbol{\theta})$ denotes the parameterized unitary that is referred to as \textit{the ansatz}, which prepares the trial state vector $\ket{\psi(\boldsymbol{\theta})}=\hat{U}(\boldsymbol{\theta})\ket{\psi_0}$. 
The performance of such approaches critically depends on the precise choice of the quantum circuit ansatz: Out of conceptual and theoretical considerations on expressivity, one cannot expect universal ans\"{a}tze to perform well for all instances of all conceivable problems.
This insight has prompted the development of a wide range of ansätze and construction strategies in recent years \cite{Cerezo_2021}. 
These range from highly chemistry-inspired ansätze, such as the \emph{unitary coupled cluster with singles and doubles} (UCCSD) 
method~\cite{romero2018} to fully problem-agnostic \emph{hardware-efficient ansätze} (HEA)~\cite{Kandala_2017}.
Additionally, ADAPT-VQE~\cite{ADAPT_VQE, Tang_2021} has been developed as an adaptive ansatz construction method that sequentially builds the circuit by adding operators with the largest energy gradient, from a predefined pool. Additionally, linear combinations of tensor-network-encoded reference states have been considered in the \emph{tensor network quantum eigensolver} (TNQE)~\cite{Leimkuhler}.
All of these methods come with their respective advantages and drawbacks, as, for example, outlined in Refs.~\cite{Cerezo_2021, 2021arXiv211105176T}. Recent works have also discussed the extent to which expressivity, trainability, and efficient classical simulation may be intertwined in variational ansätze \cite{Dequantization,PhysRevLett.131.100803}.

To date, no proven separation between quantum variational methods and classical ones has been established. And yet, this can be seen as an invitation to develop more versatile ansatz families and motivates our work to specifically address a challenge that is often overlooked: an ansatz should not only be chosen based on the problem class, but its performance can also depend sensitively on parameter variations within a single class.

This can constitute a particularly critical shortcoming, given the nature of the electronic ground state of a molecule, which typically exhibits a continuous but qualitatively varying dependence on bond distance as the system transitions between different correlation regimes. Consequently, a single fixed ansatz, even with varying parameters, will naturally exhibit variations in accuracy along the PEC. Common ansatz strategies rarely address this issue; instead, in the VQE setting, it is common to approximate the ground-state energy independently at a set of bond distances, either by using the same fixed circuit structure or by a costly optimization of an ansatz from scratch for each geometry, followed by interpolating the resulting energies across bond distances. Strictly speaking, this provides no continuous access to the wavefunction across geometries, only to interpolated energy values at non-grid points. 

Motivated by these considerations, we shift perspective and frame the computation of the PEC directly in a bond-distance-dependent setting, in which both the circuit structure and its parameters are explicitly learned as functions of molecular geometry. 
Concretely, consider a family $\mathcal{P}=\{\hat{H}(R)|R\in[R_{\min},R_{\max}]\}$ of Hamiltonians $\hat{H}(R)$, each parameterized by a bond distance $R$ between two extreme values, $R_{\min}$ and $R_{\max}$. Our approach targets the task of learning a mapping $f$ that maps bond distances to quantum circuits, i.e., 

\begin{equation}\label{eq-PQC_mapping}
    f: R \mapsto \hat{U}(R,\boldsymbol{\theta}(R)),
\end{equation}
where $\hat{U}(R,\boldsymbol{\theta}(R))$ is the unitary operator corresponding to a quantum circuit. The circuit is composed of parameterized gates, with $\boldsymbol{\theta}(R)$ being the parameters, and the function specifies the values of these parameters in terms of $R$.

Our formulation aligns with the emerging paradigm of \emph{generalizable ansätze}. Here, generalization refers to the ability to leverage knowledge of the quantum circuit and its parameters, learned from representative instances of a given Hamiltonian class, to generate quantum circuits for unseen instances drawn from the same underlying distribution of Hamiltonians, without requiring additional optimization. Similar ideas have motivated developments in the VQE context, although they have primarily focused on reusing variational parameters for new molecular geometries. For instance, Meta-VQE \cite{cervera2021meta} trains a parameterized quantum circuit on a discrete set of Hamiltonian parameters, allowing the circuit to implicitly learn a dependence of its variational parameters on the Hamiltonian. In Ref.~\cite{tao2022}, a \emph{deep neural network} (DNN) 
represents the variational parameters of the VQE as a function of the molecular geometry, trained on a discrete set of the latter. These ans\"{a}tze are usually generated through \textit{transferable} architectures. Transferability refers to the ability to carry over certain learned behavior to new domains, for example, new molecules, with little or no change to the underlying architecture. Recent examples include classical ab initio quantum chemistry, where transferability has already been demonstrated by parameterizing wavefunctions using neural networks \cite{gao2022,scherbela2021,schätzle2025, foster2025abinitiofoundationmodel}.

In this work, we learn the function $f$ by developing a transferable \emph{reinforcement learning} (RL) framework in which the agent learns 
both the circuit structure and its parameters as a function of bond distance. 

As such, our framework can be viewed as simultaneously solving many VQE problems across a continuous interval of bond distances, with the knowledge of how to construct and optimize the corresponding circuits unified within a single RL agent. Technically, this is achieved by explicitly designing the agent to enable \emph{generalization}: during training, it is exposed only to a limited, discrete set of bond distances, yet the resulting policy can generalize to arbitrary, unseen bond distances within an interval, without requiring retraining.  
The resulting ansatz can adapt to varying degrees of electron correlation across the PEC, provides access to the wavefunction at arbitrary bond distances, and has the potential to significantly reduce computational cost by identifying and reusing structural patterns across geometries. The RL framework itself is transferable to different molecular designs, without requiring the redesign of the architecture using auxiliary information.

\begin{figure*}
    \centering
    \includegraphics[width=\textwidth]{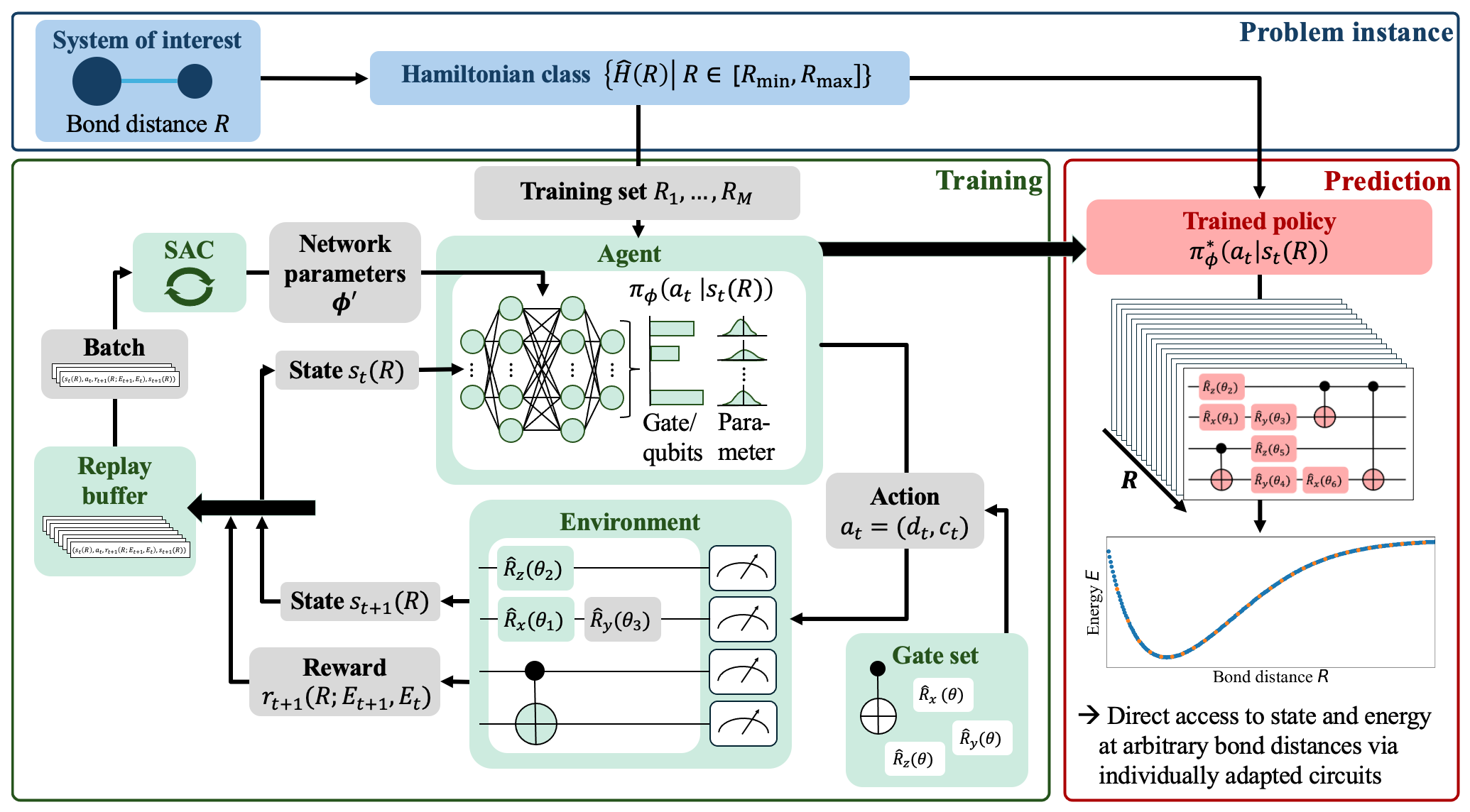}
    \caption{\textbf{Problem:} We consider the ground state energy of a molecular system, with Hamiltonian $\hat{H}(R)$, as a function of the bond distance $R\in[R_{\min},R_{\max}]$.
    \textbf{Training:} The RL agent is exposed to a discrete set $\{R_1,\dots,R_M\}$ of bond distances during training. During each episode, the agent sequentially constructs a quantum circuit from scratch by selecting gates, from a predefined gate set, and their corresponding parameters $a_t=(d_t,c_t)$. After each gate selection, the transition $(s_t(R),a_t,r_{t+1},s_{t+1}(R))$ is collected to populate the replay buffer. At regular intervals, batches are sampled from the replay buffer to update the agent’s policy network parameters $\phi$ via the Soft Actor-Critic (SAC) algorithm.
    \textbf{Prediction:} After training, the policy can predict individually adapted circuits for arbitrary bond distances $R$ within the training interval, enabling direct access to the PEC and corresponding wavefunctions.}
    \label{fig:rl_framework}
\end{figure*}

The main contributions of our work are the following.
\begin{enumerate}[topsep=0cm,left=0cm,itemsep=-0.8ex]

\item We introduce a formalized RL framework that constructs Hamiltonian-parameter-dependent quantum circuits and generates qualitatively accurate potential energy curves for molecular systems across the bond distance range, without requiring retraining for each bond distance and thereby reducing the overall training cost. See Fig.~\ref{fig:rl_framework} for a summary of the framework.

\item We apply this framework and explicitly demonstrate both the non-greedy exploration behavior of the RL agent and reduced training cost on representative examples of four- and six-qubit lithium hydride (LiH) and eight-qubit H$_4$. See Table~\ref{tab:summary_results} for a summary of the results. We also discuss how this capability can complement existing greedy or gradient-based ansatz construction methods and be integrated into hybrid design frameworks. 

\item We demonstrate that the generated quantum 
circuits reflect chemically meaningful structures and show how generalizable quantum circuit design principles can be extracted from them, providing a foundation for scalable circuit construction in larger molecular systems.
\end{enumerate}

We use RL as the basis for our framework due to its natural suitability for transfer learning, enabled by its ability to learn hierarchical feature representations that generalize across problem instances~\cite{bengio2014}. Beyond that, RL also offers distinctive characteristics that are particularly well-suited to quantum circuit design.
RL employs an inherent non-greedy optimization strategy for gate and parameter selection that does not rely on gradient information, which sets it apart from existing circuit design and optimization techniques. This enables the exploration of non-intuitive gate and parameter placements that do not yield immediate energy reductions and would likely be missed by greedy, gradient-based methods such as ADAPT-VQE \cite{Grimsley_2023}.

In addition, our RL framework operates without encoding any domain-specific or physically motivated auxiliary information: it selects gates from a hardware-efficient operator pool and learns solely through interaction with the circuit and its energy. This defines an exceptionally challenging setting for learning molecular representations, as the RL agent explores a search space that is, in principle, exponentially large with respect to circuit depth. We deliberately adopt this unconstrained setting to explore which structural patterns an RL agent can extract without prior knowledge of the system it is optimizing for. This question is particularly relevant given the limited scalability of many ML-based approaches. Demonstrating that an RL agent can extract meaningful insights from small systems may pave the way for scalable applications, where learned structures can be transferred or reused in larger settings, challenging the notion of ML as a brute-force approach dependent on large data volumes and model capacities to yield useful results. As an additional advantage, not encoding problem-specific information renders our framework readily transferable to parameter-dependent problem classes beyond quantum chemistry. For example, we showcase transferability to a job-shop scheduling problem in Appendix~\ref{app:optimization}. 

The use of RL for ground-state energy estimation with variational quantum circuits has been explored in early works~\cite{ostaszewskiReinforcementLearningOptimization2021,patelCurriculumReinforcementLearning2024,AIQT}. Beyond this, RL has been applied for constructing quantum states and for quantum circuit architecture search~\cite{Giordano_2022,zhu2023quantum}, for quantum compiling~\cite{Zhang_2020,Moro_2021,nakaji2025}, for quantum optimization and combinatorial problems~\cite{Wauters_2020,Khairy_2020,yao2020policygradientbasedquantum,fodera2024}, for reducing circuit depths~\cite{fosel2021quantum,Kundu2023EnhancingVQ}, and for decoding in quantum error correction~\cite{swekeReinforcementLearningDecoders2020}.
However, none of these works are designed for transferability on the framework level or generalization on the circuit level, nor do they leverage RL itself to optimize the variational parameters. 

Overall, our work provides the starting point for using RL to expand the current toolbox for ansatz design, due to its potential for transferability, non-greediness, and the ability to construct interpretable and generalizable circuits.

\section{Framework}\label{framework}

\subsection{Problem formulation}

Our goal is to determine the potential energy curve for a Hamiltonian family $\mathcal{P} = \{\hat{H}(R)| R\in [R_{\min}, R_{\max} ]\}$, where $\hat{H}(R)$ is the molecular system Hamiltonian as a function of the bond distance $R$. 
The potential energy curve, denoted as the function $R\mapsto E_{\text{GS}}(R)$, is defined as the lowest eigenvalue of the Hamiltonian $\hat{H}(R)$, with the corresponding ground state vector $\ket{\psi_{\text{GS}}(R)}$ satisfying
\begin{align}
    \hat{H}(R)\ket{\psi_{\text{GS}}(R)}=E_{\text{GS}}(R)\ket{\psi_{\text{GS}}(R)}.
\end{align}

The objective of our RL method is then to learn a quantum circuit $\hat{U}(R;\boldsymbol{\theta}(R))$ and the corresponding optimal parameters $\boldsymbol{\theta}$ as an $R$-dependent function, preparing a trial state vector
\begin{align}
    \ket{\psi(R;\boldsymbol{\theta}(R))}&=\hat{U}(R;\boldsymbol{\theta}(R))\ket{\psi_0},
\end{align}
with a fixed initial state vector $\ket{\psi_0}$, for $R\in[R_{\min},R_{\max}]$. After the RL is trained, the potential energy curve can be directly deduced by
\begin{align}
    E(R)&= \braket{\psi(R;\boldsymbol{\theta}(R))|\hat{H}(R)|\psi(R;\boldsymbol{\theta}(R))},\\
    &\qquad\qquad\qquad\qquad\qquad\forall R\in [R_{\min}, R_{\max} ].\notag
\end{align}

\subsection{Theory}
\label{Theory}
We formulate our reinforcement learning approach to constructing an $R$-dependent quantum circuit and selecting its corresponding parameters within the framework of \emph{Markov decision processes (MDPs)}~\cite{Put14_book,Sutton2018}. An overview of the overall framework, including the problem instance, the training procedure, and the prediction, is provided in Fig.~\ref{fig:rl_framework}.
An MDP is defined by the 4-tuple $\langle\mathcal{S},\,\mathcal{A},\,\mathcal{R},\,p(s_{t+1},r_{t+1}|s_{t},a_t)\rangle$, where $\mathcal{S}$ represents the set of states, $\mathcal{A}$ the set of actions, $\mathcal{R}$ the reward function and $p(s_{t+1},r_{t+1}|s_{t},a_t)$ the transition probabilities between states and the associated reward \cite{Sutton2018}. To enable transferable learning, the bond distance $R$ is embedded into the state representation $s_t(R)$. As illustrated in Fig.~\ref{fig:rl_framework}, the agent selects an action $a_t \in \mathcal{A}$  based on the observation of the $R$-dependent state $s_t(R) \in \mathcal{S}$ at each discrete time step $t$ of the episode, $t\in\{0,1,2,\dotsc,T-1\}$. Actions are selected according to a policy $\pi(a_t|s_t(R))$, which maps each 
state $s_t(R)$ to a probability distribution over all possible actions $a_t$. 
The chosen action $a_t$ triggers a state transition from $s_t(R)$ to $s_{t+1}(R)$ according to the transition probability $p(s_{t+1}(R),r_{t+1}|s_t(R), a_t)$.
Based on the chosen action, and on the new state $s_{t+1}(R)$, the agent receives a reward $r_{t+1}\in \mathcal{R}$. 
The objective of the RL agent is to learn the policy $\pi^*(a_t|s_t(R))$ that maximizes the cumulative sum of rewards over one episode, i.e.,
\begin{align}
        \pi^* = \arg\max_{\pi} \mathbbm{E}_{\pi}\bigg[\sum^{T-1}_{k=0}r_{k+1}\bigg].\label{optpolicy}
\end{align}
In our work, we employ the \emph{soft actor-critic (SAC) algorithm} \cite{haarnoja2018, haarnoja2019}, generalized to discrete-continuous action spaces \cite{Erdman_2023}, to jointly learn action selection $a_t=(d_t,c_t)$, where $d_t$ denotes the discrete gate selection and $c_t$ their associated continuous parameter. The policy $\pi_{\phi}(a_t|s_t(R))$ is parameterized by a neural network with parameters~$\phi$.
SAC is based on an entropy-regularized RL objective that promotes exploration and improves sample efficiency compared to other state-of-the-art algorithms \cite{haarnoja2018}. 
During training, state-action transitions of the form $(s_t(R),a_t,r_{t+1},s_{t+1}(R))$ are stored in a replay buffer and periodically sampled to update and continuously refine the policy. 
Generalization across different bond distances $R$ is achieved by exposing the agent to a discrete training set ${ R_1, R_2, \dots ,R_M }$, such that the policy learns to condition its action choices on the desired value of $R$. A detailed description of the SAC algorithm and the corresponding learning routine is presented in Appendix~\ref{SAC}. For our purpose of constructing quantum circuit architectures, we define the 4-tuple of the corresponding MDP as follows

\textbf{States $\mathcal{S}$}. 

The set of quantum states prepared by the agent-constructed circuit on an $n$-qubit system is
\begin{align}
\mathcal{S}_{\psi} = \big\{\, 
  &|\psi(t,R;\boldsymbol{\theta})\rangle = \hat{U}(t,R;\boldsymbol{\theta})|\psi_0\rangle\,\big| \\
  & \boldsymbol{\theta} \in (-\pi,\pi],\, R \in [R_{\min},R_{\max}],\, t \in [0,T] 
\,\big\} \subset \mathbb{C}^{2^n}.\nonumber
\end{align}

In our RL formulation, the state space of the MDP is given by a classical representation of the physical states 
$\ket{\psi(t,R,\boldsymbol{\theta})} 
\in \mathcal{S}_{\psi}$. 
Each quantum state is 
mapped 
to a real-valued, normalized vector, and 
augmented with the current 
step $t$ and a feature 
encoding 
of $R$, 
\begin{align}
    s_t(R) \coloneqq \Big(
        \text{Re}(\ket{\psi(t, R; \boldsymbol{\theta})}),\,
        \text{Im}(\ket{\psi(t, R; \boldsymbol{\theta})}),\, \notag \\
        \tfrac{t}{T - 1},\, g(R)
    \Big)
    \in \mathbbm{R}^{2^{n+1} + 1 + J},
    \label{RLstate}
\end{align}
where $t/(T-1)$ normalizes the current step $t$ with respect to the total number of steps $T$, starting from $t=0$. 
The Gaussian featurization $g(R): \mathbbm{R}\rightarrow \mathbbm{R}^J$, as detailed in Appendix~\ref{app:state_representation}, maps $R$ into a $J$-dimensional feature space using Gaussian kernels, which enhances robustness and generalization in learning~\cite{Hermann_2020}. Additional details on the 
state representation are provided in Appendix~\ref{app:state_representation}.

\textbf{Actions $\mathcal{A}$.}
Each action specifies the placement of a quantum gate on specific qubits, including the selection of its gate parameter. 
These quantum gates can be selected from the gate set 
\begin{align}
        \mathcal{G} = \left\{ \mathrm{CNOT},\, \hat{R}_x(\theta),\, \hat{R}_y(\theta),\, \hat{R}_z(\theta) \;\middle|\; \theta \in (-\pi, \pi] \right\},\label{operator_pool}
\end{align}
where CNOT is a two-qubit entanglement gate, and  $\hat{R}(\boldsymbol{\theta})$ are single-qubit rotation gates.
This results in a discrete set of gate–placement choices $A$ with cardinality $\mathcal{G}$ is $|A|=2\binom{n}{2}+3n$, with the first term representing the number of possible combinations of the target and control qubits for the CNOT gate, and the second term accounting for the single-qubit rotation gates applied to each qubit. 
When applied within the RL framework, actions are represented as a tuple $a_t=(d_t,c_t)$, where the discrete actions $d_t\in\{1, \dots, |A|\}$ determine the gate type and its target qubits through an indexed array encoding, as detailed in Appendix~\ref{app:action_encoding}. 
The continuous action $c_t\in (-\pi,\pi]$ specifies the corresponding gate parameter, conditioned on the choice of $d_t$.

\textbf{Transition probabilities $p(s_{t+1}(R),r_{t+1}|s_t(R),a_t)$.} 
The transition probabilities, whose explicit knowledge is not required in the SAC algorithm, capture the probability of the environment changing from $s_t(R)\rightarrow s_{t+1}(R)$ and of receiving the reward $r_{t+1}$ given the action $a_t$. Given our choice of state, action, and reward spaces, the transition probabilities are fully deterministic, guaranteeing that our problem formulation is indeed Markovian.
 
\textbf{Reward $\mathcal{R}$.} The main objective of the RL agent is to sequentially select $T$ gates from the gate set $\mathcal{G}$ to construct a quantum circuit that minimizes the final energy $E_T$. 
This is achieved through a reward function of the form
\begin{align}
    r_{t+1}(R;E_{t+1},E_t)=f(R;E_{t+1})-f(R;E_t),
\end{align}
where $f$ is a strictly monotonic decreasing function, to ensure lower energies are rewarded more. 
As a result, the return $G$ can be expressed as 
\begin{equation}
    G\coloneqq\mathbbm{E} \left[\sum_{t=0}^{T-1}r_{t+1}\right]=\mathbbm{E}[f(R;E_T)]-\mathbbm{E}[f(R;E_0)].\label{return}
\end{equation}
Since $\mathbbm{E}[f(E_0)]$ is a constant, and given the properties of $f(E)$, maximizing the return $G$ is equivalent to minimizing the final energy $E_T$ without any constraints on intermediate steps. This highlights the inherently non-greedy nature of the RL approach.

Within our framework, the reward function is explicitly defined as

\begin{align} f(R;E_{t})&=c_{\exp}f_{\text{exp}}(R;E_{t})+c_{\text{lin}} f_{\text{lin}}(E_{t}),\label{reward}
\end{align}
which consists of an exponential and a linear component, weighted by coefficients $c_{\exp}$ and $c_{\text{lin}}$, respectively.
The exponential component serves as the primary driver of the reward signal and is defined by

\begin{align}
    f_{\text{exp}}(R;E_{t}) = \exp\left(-\frac{E_{t}-\mu_R}{\sigma_R}\right),
\end{align}
where $\mu_R$ and $\sigma_R$ are dynamically updated during training to shift the center and slope of the exponential function towards the agent's lowest observed energy values from previous training episodes. 
Inspired by curriculum learning \cite{narvekar2020curriculum}, this dynamic adjustment ensures that the agent's achieved energy is always evaluated relative to its lowest-achieved energy values so far, effectively pushing the agent to continuously improve. This mechanism is particularly important in our quantum chemistry application, where accurately capturing the correlation energy requires an extremely sensitive energy scale to be appropriately rewarded. The dynamic exponential function is thus used to amplify energy changes in this otherwise small energy region. Note that the dynamic adaptation of the reward introduces certain technical subtleties, which are simplified in the main text but discussed in more detail in Appendix~\ref{app:reward_design_details}. The
linear reward term, given by $f_{\text{lin}}(E_t)=-E_t$, provides only a minor stabilizing contribution and is also described in more detail in Appendix~\ref{app:reward_design_details}.

\section{Experimental setup}\label{experiments}

\subsection{Quantum chemistry} \label{Quantumchem}
We evaluate our framework on the task of determining the potential energy curve for two molecular systems: two instances of \emph{Lithium Hydride} (LiH) and a \emph{four-atom hydrogen chain} (H$_4$).
As a preliminary step toward the PEC simulations, we first use our framework to compute the ground-state energy at a single bond distance for each molecule.
For all molecules, we have considered the vacuum state vector  $\ket{0}^{\otimes n}$, the \emph{Hartree–Fock} 
(HF) state, and an arbitrarily chosen state as an initial state, and ultimately selected the one that yielded the best average performance. A  detailed description of the molecular setups is provided in Appendix~\ref{app:quantumchem}.

\paragraph{Lithium hydride (LiH).} The qubit Hamiltonians for LiH are computed in the STO-3G basis, using an active space that freezes the core and removes the weakly correlated $p_x$ and $p_y$ orbitals, leading to an active space of two electrons and three spatial orbitals. Whenever we refer to the FCI energy, we mean the exact ground state energy of this active space. 
The first instance we consider, the four-qubit LiH, is constructed by mapping the fermionic Hamiltonian of LiH to a four-qubit Hamiltonian using the parity mapping~\cite{Seeley_2012}. The second instance we consider, the six-qubit LiH, is generated by mapping the fermionic Hamiltonian to a six-qubit Hamiltonian using the Jordan-Wigner mapping~\cite{Seeley_2012}.
Four-qubit LiH and six-qubit LiH are examined at a fixed bond distance of 2.2 \AA. The potential energy curves are evaluated over a bond distance range $1.0$\,\AA--$4.0$\,\AA\,.

\paragraph{Four-atom hydrogen chain (H$_4$).} The Hamiltonian for H$_4$ is computed in the STO-3G basis using the Jordan-Wigner mapping~\cite{Seeley_2012}, which results in an eight-qubit Hamiltonian. Additionally, the orbitals are optimized as described in Ref.~\cite{Kottmann_2023}, enabling the separable pair approximation (SPA) ansatz to yield accurate energy estimates. 
Due to its exceptionally low gate count of $10$, the SPA ansatz provides a useful benchmark for assessing the efficiency of the circuits generated by our RL agent. H$_4$ is first examined at a fixed bond distance of 1.5\,\AA. The potential energy curve of H$_4$ is evaluated over the restricted bond distance range 0.5--1.55~\AA. At larger bond distances, the spectrum becomes increasingly near-degenerate; in this range, the RL agent is prone to converge to energetically favorable yet symmetry-broken states (as also reported in Ref.~\cite{bertels2022}), and the resulting circuit structures are not informative for identifying physically meaningful structures.

\subsection{Training, implementation and evaluation}
Since RL training is stochastic by design, each molecular system is evaluated through twelve independent training runs. 
The agent's performance depends on several hyper-parameters, as listed in Appendix~\ref{app:hyper-parameters}, which were manually tuned based on preliminary simulations. Since the HEA pool already defines a large and expressive search space, increasing the maximum gate count would rapidly increase the number of possible circuit configurations and the problem's complexity. Thus, we intentionally restricted the maximum number of gates to a narrow range (approximately $12$ gates) per setup.  
Training is performed on bond distances sampled from a discretization of the bond distance interval in steps of $0.1$\,\AA. In SAC, actions are sampled stochastically via $a_t \sim \pi(a_t | s_t(R))$, with the resulting transition tuple $(s_t(R), a_t, r_{t+1}, s_{t+1}(R))$ stored in the replay buffer. We refer to these as non-evaluation episodes. To assess the agent’s learning progress, every tenth episode is instead sampled deterministically using $a_t = \arg\max_a \pi(a | s_t(R))$, which we refer to as an evaluation episode. After training, the learned policy is used to predict circuits for the same bond distance interval, discretized with a finer step size of 0.01 \AA.
For each setup, we evaluate the average and best-case error relative to the FCI energy.  All values are reported with an asymmetric standard deviation of the form 
$\bar{x}^{\sigma_{+}}_{\sigma_{-}}$, to prevent violations of the variational principle that could arise from using symmetric standard deviations. The statistical evaluation of the PECs is described in detail in the Appendix~\ref{app:statistical_methods}.

\section{Results}\label{results}

\begin{table*}[t!]
\centering
\begin{tabular}{lccccccc}
\toprule
\setlength{\tabcolsep}{5pt} 
& \multicolumn{3}{c}{Single bond distance} 
& \multicolumn{4}{c}{Potential energy curve} \\
\cmidrule(lr){2-4}\cmidrule(lr){5-8}
System 
& Avg.\ err.\ (Ha) & Best err.\ (Ha) & Cost
& Avg.\ err.\ (Ha) & Best err.\ (Ha) & Avg.\ err.\ at ref.\ $R$ (Ha) & Cost \\
\midrule
Four-qubit LiH     
  & 0.0041  & 0.0007  & 30{,}000  
  & 0.0136  & 0.0058  & 0.0064 & 1{,}380  \\
Six-qubit LiH      
  & 0.0110  & 0.0078  & 40{,}000  
  & 0.0161  & 0.0159  & 0.0124 & 1{,}724  \\
Eight-qubit H$_4$  
  & 0.1463  & 0.0948 & 40{,}000  
  & 0.0815  & 0.0538  & 0.1594 & 4{,}545  \\
\bottomrule
\end{tabular}
\caption{ Summary of our results. 
\emph{Avg.\ err.} and \emph{Best err.} denote the mean and minimum absolute deviation from the
FCI reference energy in the single-bond setting and from the FCI reference PEC in the PEC setting,
respectively, each computed across 12 independently trained runs (12 single-bond runs and 12 PECs).  \emph{Avg.\ err.\ at ref.\ $R$} denotes the mean absolute deviation of the PEC from the FCI
reference at the bond distance that is also used in the single-bond setting
(2.2\,\text{\AA} for LiH and 1.5\,\text{\AA} for H$_4$). This provides a direct accuracy
comparison with the single-bond results. The cost per bond distance denotes
the average number of RL episodes per bond length. See Appendix~\ref{app:statistical_methods} for
details of the calculations.}
\label{tab:summary_results}
\end{table*}

Table~\ref{tab:summary_results} provides an overview of the performance of our RL-based framework across all molecular systems considered. For each molecule, we report the average and best-case errors with respect to the FCI reference, as well as the average training cost per bond distance for the single-bond-distance and PEC experiments. For the PEC setting, we additionally report the average error of the PEC at the bond distance used in the single-bond-distance setup for that molecule, to enable a direct accuracy comparison between the two tasks. In the following, we discuss the individual experiments in more detail to provide additional context and qualitative insight.

\subsection{Four-qubit LiH}

\subsubsection{Single bond distance for four-qubit LiH}
\label{SBDLiH4}

\begin{figure}[t]
    \centering
\includegraphics[width=\columnwidth]{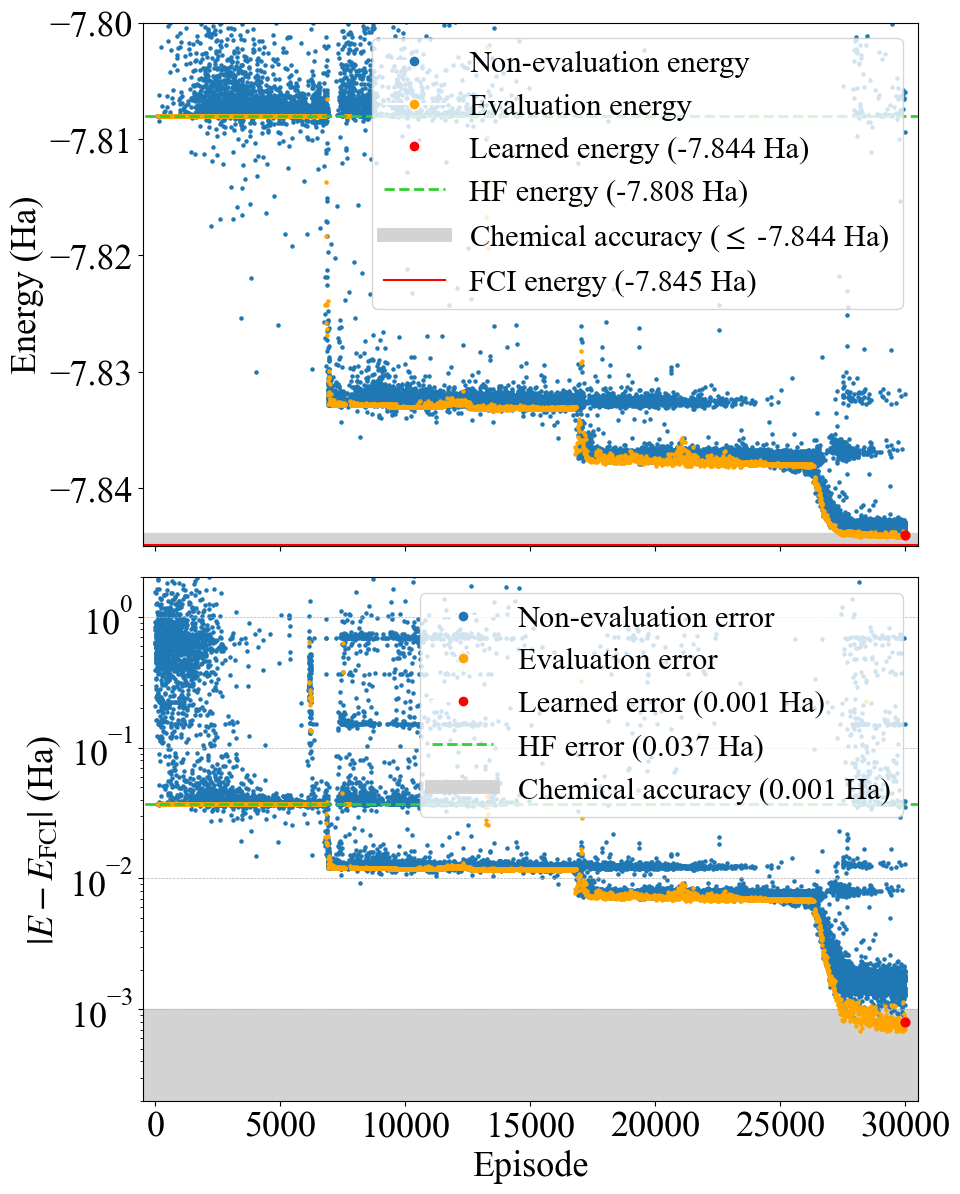}
    \caption{
        Progression of the obtained energies (top) and errors (bottom) during training of one agent for four-qubit LiH at a fixed bond distance of 2.2~\AA. 
        Each point represents the final energy and error of one episode. 
        Blue points indicate non-evaluation energies,
        while orange points represent evaluation energies. 
        The dashed green line denotes the HF energy (top) and the HF error (bottom). 
        The red line in the top plot represents the FCI energy. 
        The gray area marks the region within chemical accuracy. 
        The final learned energy and error are marked by a red point and are $-7.8441$~Ha and $0.0007$~Ha, respectively. 
        The corresponding hyper-parameters are listed in Appendix~\ref{app:hyper-parameters}. }
    \label{fig:LiH4_energy}
\end{figure}

Our framework converged to energies within chemical accuracy (0.001 Ha) in five of the twelve evaluated runs for four-qubit LiH at the single-bond distance of 2.2 \AA.
The corresponding hyper-parameters are specified in Appendix~\ref{app:hyper-parameters}.
The mean learned energy across all runs is 
$-7.8408^{+0.0046}_{-0.0030}$ Ha, deviating from the FCI energy by 0.0041 Ha.
Comprehensive statistical analyses of all 12 runs are presented in Appendix~\ref{app:LiH4_supplementary_material}. 
In the following, we discuss key aspects, such as convergence behavior and the resulting circuit construction, based on one representative best-case run out of 12 runs, as shown in Fig.~\ref{fig:LiH4_energy}. The upper plot of Fig.~\ref{fig:LiH4_energy} illustrates the energies obtained during training, while the lower plot shows the error with respect to the FCI energy on a logarithmic scale. 
The energies and errors shown in blue correspond to non-evaluation episodes generated by stochastic policy sampling, while the orange points represent evaluation results obtained via deterministic policy selection.
The red point marks the final deterministically selected episode and thus corresponds to the final learned energy of $-7.8441$ Ha, which achieves chemical accuracy with an error of $0.0007$ Ha.

Most prominently, Fig.~\ref{fig:LiH4_energy} illustrates the agent's stepwise learning behavior, characterized by repeated stabilization at intermediate energy plateaus, before successfully escaping towards lower energy levels.
These energy plateaus reflect the agent’s stepwise inclusion of energetically impactful basis states on top of the HF reference, leading to a systematic increase in electronic correlation. Chemically, this is most naturally compared to an UpCCD description~\cite{Elfving2021}: the first post-HF energy plateau is reproduced by UpCCD with paired doubles only, whereas subsequent plateaus are consistent with UpCCD, including a sparse subset of additional contributions.

Importantly, these plateaus represent robust local minima within the current parameterization, so escaping them is typically not possible by small local adjustments alone. The agent's ability to identify and escape these stable configurations without becoming trapped highlights its effective use of exploration strategy and non-greedy decision-making.

\begin{figure}[t!]
\includegraphics[width=0.95\columnwidth]{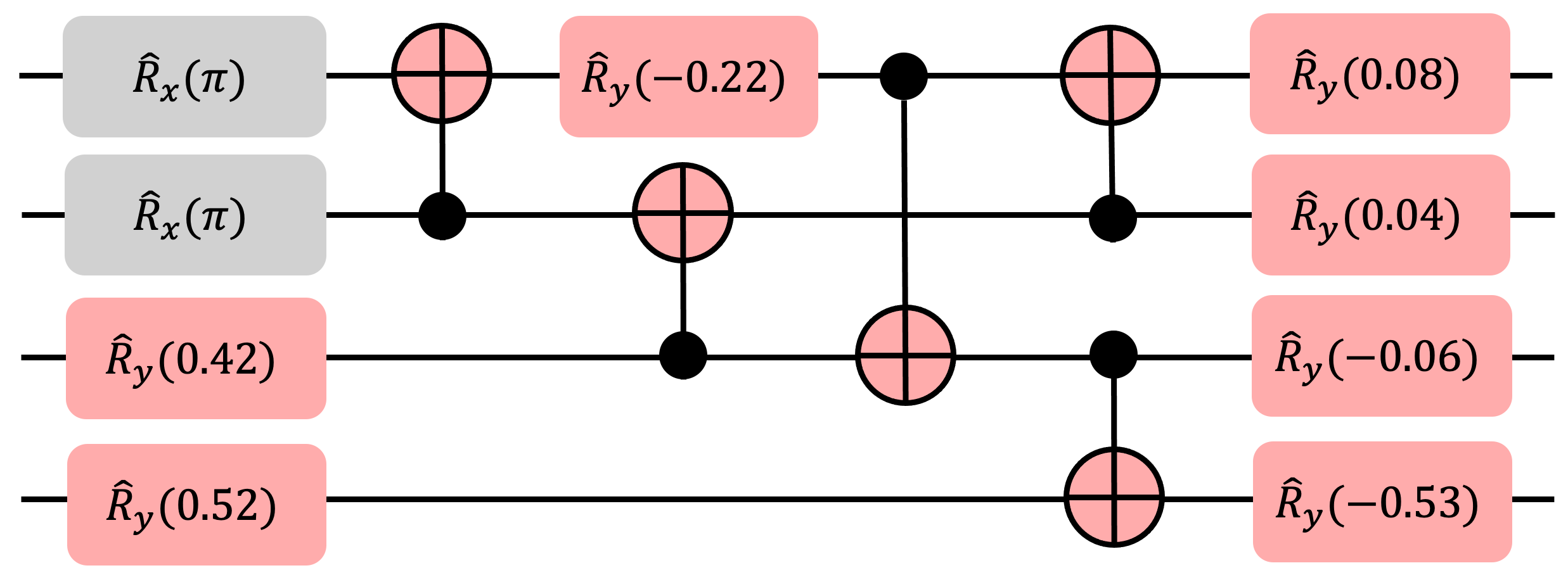}
\caption{Learned circuit corresponding to the final learned energy of the four-qubit LiH molecule shown in Fig.~\ref{fig:LiH4_energy}. 
The circuit has a depth of six and consists of 12 gates, selected by the agent, with the first two grey-colored gates predefined to initialize the system in the HF state.}
\label{LiH4_circuit}
\centering
\end{figure}

Fig.~\ref{LiH4_circuit} illustrates the corresponding circuit learned by the RL agent. 
The first two $\hat{R}_x(\pi)$ gates of the circuit are 
pre-defined to initialize the system in the HF state. 
The circuit has a depth of six (sequential gate layers that can be executed in parallel), consists of five CNOT gates, and includes seven rotational parameters. 
Its depth could be reduced to five by removing the first $R_x(\pi)$ and CNOT gate. Since the Hamiltonian is already compressed through the parity encoding (see also the discussion on the six-qubit system), we can assume that all qubits need to be correlated, leading to a depth of four as a lower bound to the optimal depth. Accordingly, the RL agent’s identified circuit is notably shallow and near-optimal in depth for this case.

\begin{figure}[t!]
\includegraphics[width=0.95\columnwidth]{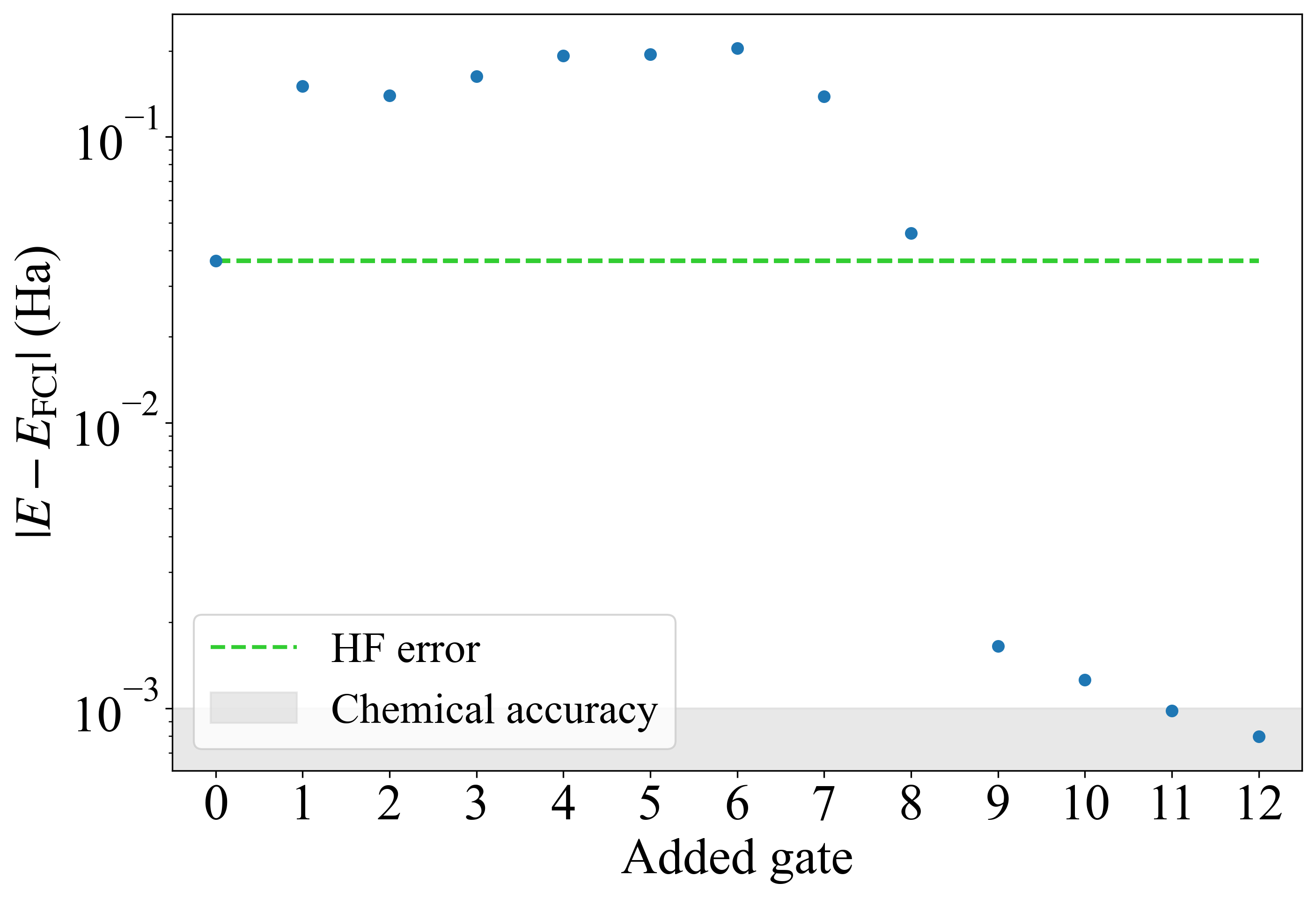}
\caption{Energy error relative to the FCI energy per added gate of the learned circuit shown in Fig.~\ref{LiH4_circuit} for the four-qubit LiH molecule.
}
\label{LiH4_nongreedy}
\centering
\end{figure}

Fig.~\ref{LiH4_nongreedy} shows how the energy error changed with each sequentially added gate by the agent to construct the circuit in Fig.~\ref{LiH4_circuit}.
As shown, the agent's selected gate sequence results in a non-monotonic decrease in energy per added gate, with several gates initially increasing energy. 
This reflects how the RL agent deliberately leverages its characteristic non-greedy, long-term strategy—a distinctive imprint of RL-based circuit design—to achieve a shallow and chemically accurate circuit.

\subsubsection{Potential energy curve for four-qubit LiH}
\label{PECLiH4}
Across 12 independent runs, our framework generated the PEC for four-qubit LiH over the bond distance interval 1--4~\AA, with an average error of $0.0136^{+0.0156}_{-0.0096}$ Ha relative to the FCI 
energy. This corresponds to an average performance approximately $5.1$ times more accurate than that of the HF approximation for the PEC.
A comprehensive statistical analysis of all twelve runs is provided in Appendix~\ref{app:LiH4_supplementary_material}, with the corresponding hyper-parameters listed in Appendix~\ref{app:hyper-parameters}.

\begin{figure}[t!]
    \centering
    \includegraphics[width=0.95\columnwidth]{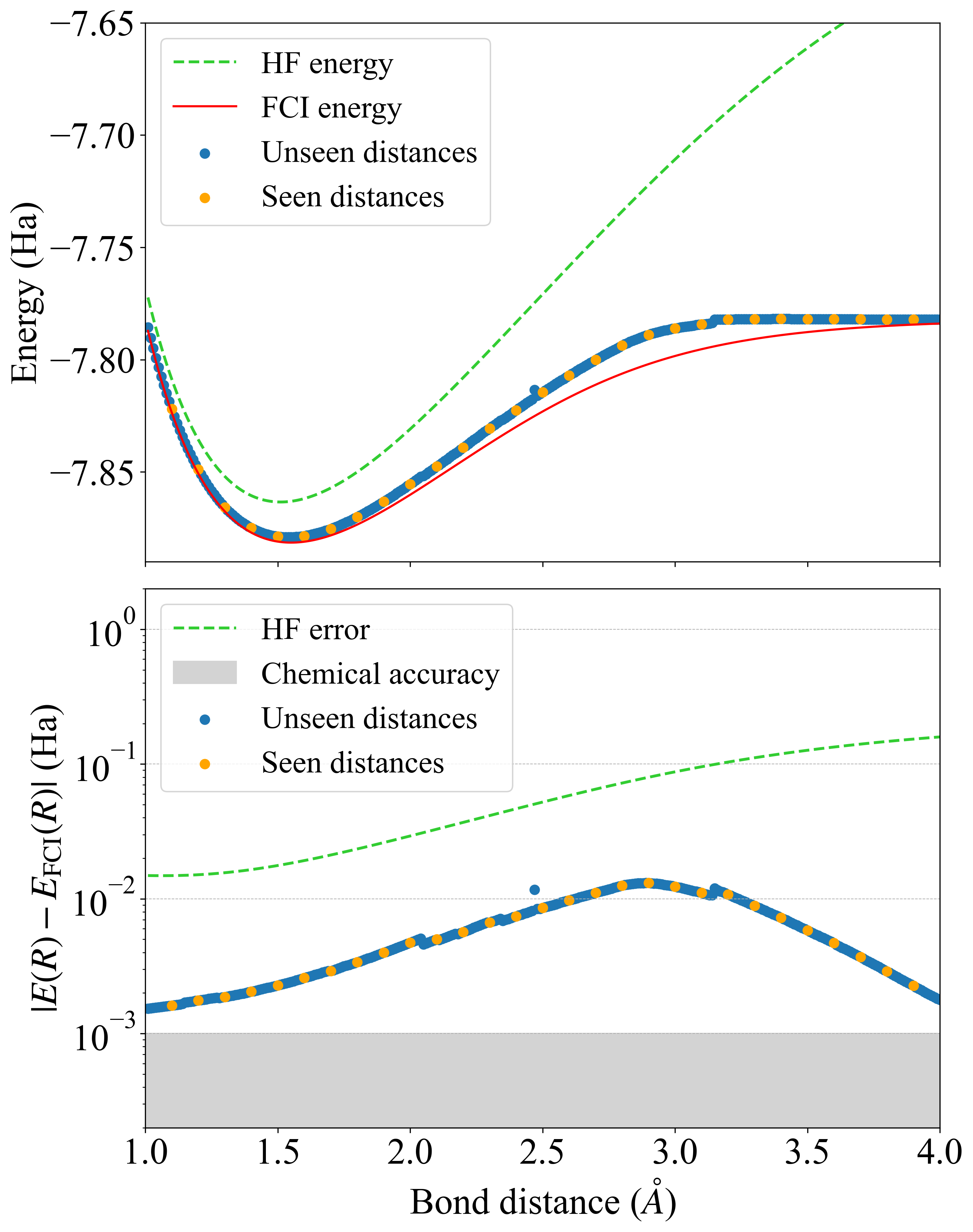}
\caption{Potential energy curve for four-qubit LiH (top) and absolute energy differences with respect to the FCI energy (bottom). 
The orange energies and errors correspond to bond distances that were seen during training, while the blue energies and errors indicate predictions for unseen bond distances. The red curve represents the exact potential energy curve for the four-qubit LiH system, and the gray area indicates the region of chemical accuracy.  
The green curve represents the HF approximation of the potential energy curve. 
The hyper-parameters are listed in Appendix~\ref{app:hyper-parameters}.}
\label{LiH4_Transferability}
\end{figure}

In the following, we analyze the performance and the circuit architecture generated by our framework based on one representative best-case run out of the twelve runs, shown in Fig.~\ref{LiH4_Transferability}. The upper panel of Fig.~\ref{LiH4_Transferability} depicts the PEC across the bond distance range 1--4~\AA, whereas the lower panel shows the corresponding error relative to the FCI energy (red curve). The orange points represent the learned energies and errors at the bond distances $R$ that were used during training to condition the circuit on these $R$ values. The blue energies and errors correspond to predictions obtained by generalization of the model. 
For this representative run, the average error over the entire PEC is $0.0058^{+0.0045}_{-0.0030}$ Ha, which is approximately twelve times more accurate than the HF approximation (dashed green line). 
The generated PEC provides a significantly improved qualitative approximation of the overall FCI curve compared to the HF method. This improvement can be attributed to the ability of our approach to adapt to different correlation regions of the PEC, whereas the HF method is primarily tailored to describe the molecular region but becomes increasingly inaccurate as the dissociation limit is approached. 
Notably, the PEC error increases toward $2.8$\,\AA\, before decreasing again, marking the transition from the molecular regime to the dissociation limit. Difficulties in maintaining accuracy in this intermediate region have also been qualitatively observed in PECs generated with chemically motivated ansätze~\cite{Kottmann_2024,burton2022}. In our case, this behavior can be attributed to the RL agent transitioning from a symmetry-preserving to a symmetry-broken solution to maintain an accurate energy estimate; see Appendix~\ref{app:LiH4_supplementary_material} for supporting details. 

From a chemistry perspective, this behavior exemplifies the Löwdin symmetry dilemma, which states that enforcing symmetry in HF calculations can yield higher energies than allowing symmetry-breaking~\cite{bertels2022,Lykos1963}. Similar observations have also been reported in the context of VQE~\cite{bertels2022}. The issue becomes particularly relevant at larger bond distances, where weakening of interatomic interactions increases degeneracy, thereby raising the likelihood of convergence to energetically favorable states that break the true ground state's symmetries.
From an RL perspective, the agent’s exploitation of symmetry breaking to achieve accurate ground-state energies can be interpreted as a loophole in the reward function — somewhat analogous to how RL agents in video games discover unintended strategies in the environment to maximize reward~\cite{Lehman2020}. To avoid convergence to physically invalid solutions, this observation suggests that incorporating symmetry constraints into the reward function is essential.

Regarding the training effort of the PEC generation compared to single-distance training: The PEC was trained on 29 bond distances using $40{,}000$ episodes, whereas the single-distance setup at $2.2$\,\AA, shown in Fig.~\ref{fig:LiH4_energy}, was trained using $30{,}000$ episodes.
Thus, the transferable setup effectively reduces the required training effort to approximately $1{,}380$ episodes per bond distance, representing a 22-fold improvement in training efficiency compared to the single-bond-distance setup. This efficiency comes with a modest accuracy cost: The resulting average error across all PEC runs is $0.0136^{+0.0156}_{-0.0096}$ Ha compared to the average error observed in the single-distance training setup of $0.0041^{+0.0046}_{-0.0030}$ Ha. Despite this accuracy trade-off, this highlights the RL agent’s ability to efficiently generalize both energy estimates and the corresponding wave functions across the continuous bond-distance interval of 1--4~\AA, substantially expanding the learned solution space at a very low additional training cost.

\begin{figure}[t!]
    \centering
   \includegraphics[width=0.95\columnwidth]{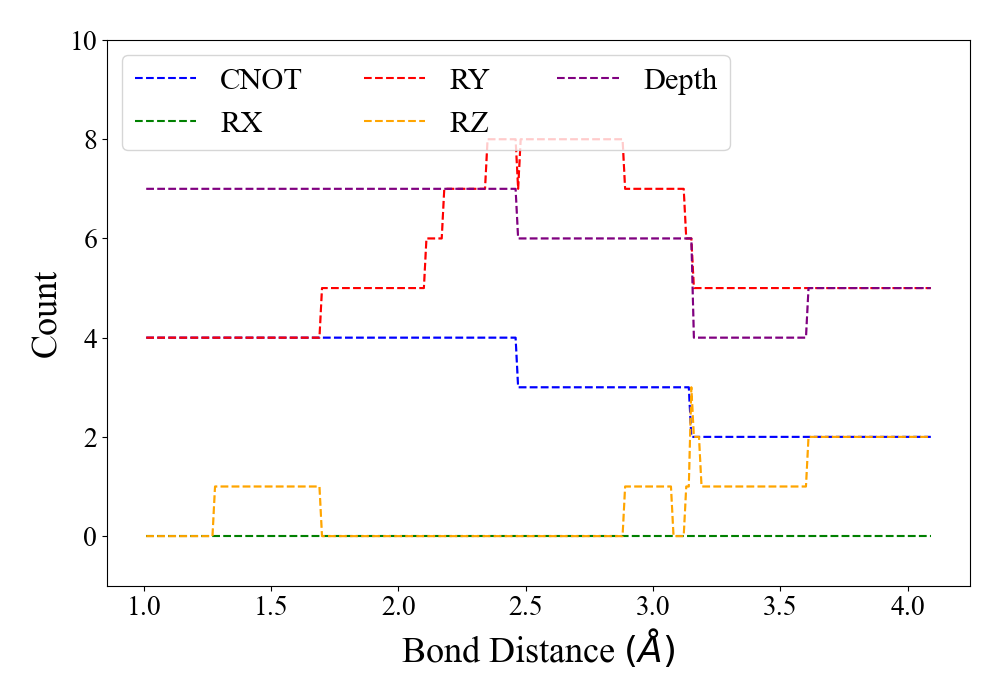}
    \caption{Count of specific operators CNOT (blue), $\hat{R}_x$ (green), $\hat{R}_x$ (red), $\hat{R}_z$ (yellow) and the total circuit depth (purple) over a bond distance range of 1.0~\AA\, and 4.0~\AA\, for the four-qubit LiH molecule for a single run. }
    \label{topology_LiH4}
\end{figure}

The large dataset of circuits generated by the agent across the PEC allows us to analyze
circuit structures across bond distances and whether their structure and gate composition reflect \textit{physical and chemical insights} about the underlying molecular system. For this purpose, we preprocessed the circuits constructed by the agent to remove trivial redundancies (see Appendix~\ref{circuits_pre-processing}). 
Fig.~\ref{topology_LiH4} illustrates how the circuit topology evolves with bond distance, showing the gate counts for CNOT (blue), $\hat{R}_x$ (green), $\hat{R}_y$ (red), $\hat{R}_z$ (yellow), and the total circuit depth (purple) over the range from 1.0 to 4.0\,\AA. Notably, the agent makes no use of $\hat{R}_x$ gates, employs $\hat{R}_z$ gates only sparsely, and consistently applies a large number of $\hat{R}_y$ rotations across all bond distances, with their number further increasing as the bond distance grows. This reflects the fact that the ground-state wavefunction of molecular systems is real-valued, and the use of $\hat{R}_x$ gates would unnecessarily introduce complex amplitudes. In contrast, $\hat{R}_y$ gates generate the required real-valued superpositions, while any necessary phase adjustments can be handled by $\hat{R}_z$ gates. The increase of $\hat{R}_y$ gates with bond distance suggests that the agent adapts to the growing electronic correlation by increasing the number of local superpositions in the ansatz. Thus, the agent implicitly learned that the operator pool can be reduced to just $\hat{R}_y$, $\hat{R}_z$, and CNOT gates, exactly the building blocks needed to construct chemically motivated ansätze, where single and double excitations can be compiled into standard gate sequences involving these gates~\cite{Arrazola_2022}. The observed reduction in circuit depth, CNOT count, and $\hat{R}_y$ usage beyond $3.0$~\AA, is likely a consequence of the previously discussed symmetry breaking. A detailed visualization of the circuit’s evolution with bond distance is provided in Appendix~\ref{app:LiH4_supplementary_material}. In particular, this reveals that the agent learns a stable core entanglement pattern that is only slightly adapted and gradually enriched with additional $\hat{R}_y$ gates as the bond distance increases, while the circuit parameters are continuously adjusted.

\subsection{Six-qubit LiH} \label{LiH6_transferability}
Across 12 independent runs, the generated PECs for six-qubit LiH over the bond distance range 1--4~\AA\ yield a mean error of $0.0161^{+0.0136}_{-0.0036}$ Ha relative to the FCI energy. On average, this corresponds to an accuracy approximately $4.3$ times better than the HF approximation, and slightly below the average performance achieved for the four-qubit LiH PEC. Details of the twelve individual runs are provided in the Appendix~\ref{app:LiH6_supplementary_material}, with their corresponding hyper-parameters listed in Appendix~\ref{app:hyper-parameters}.

\begin{figure}
    \centering
        \includegraphics[width=0.95\columnwidth]{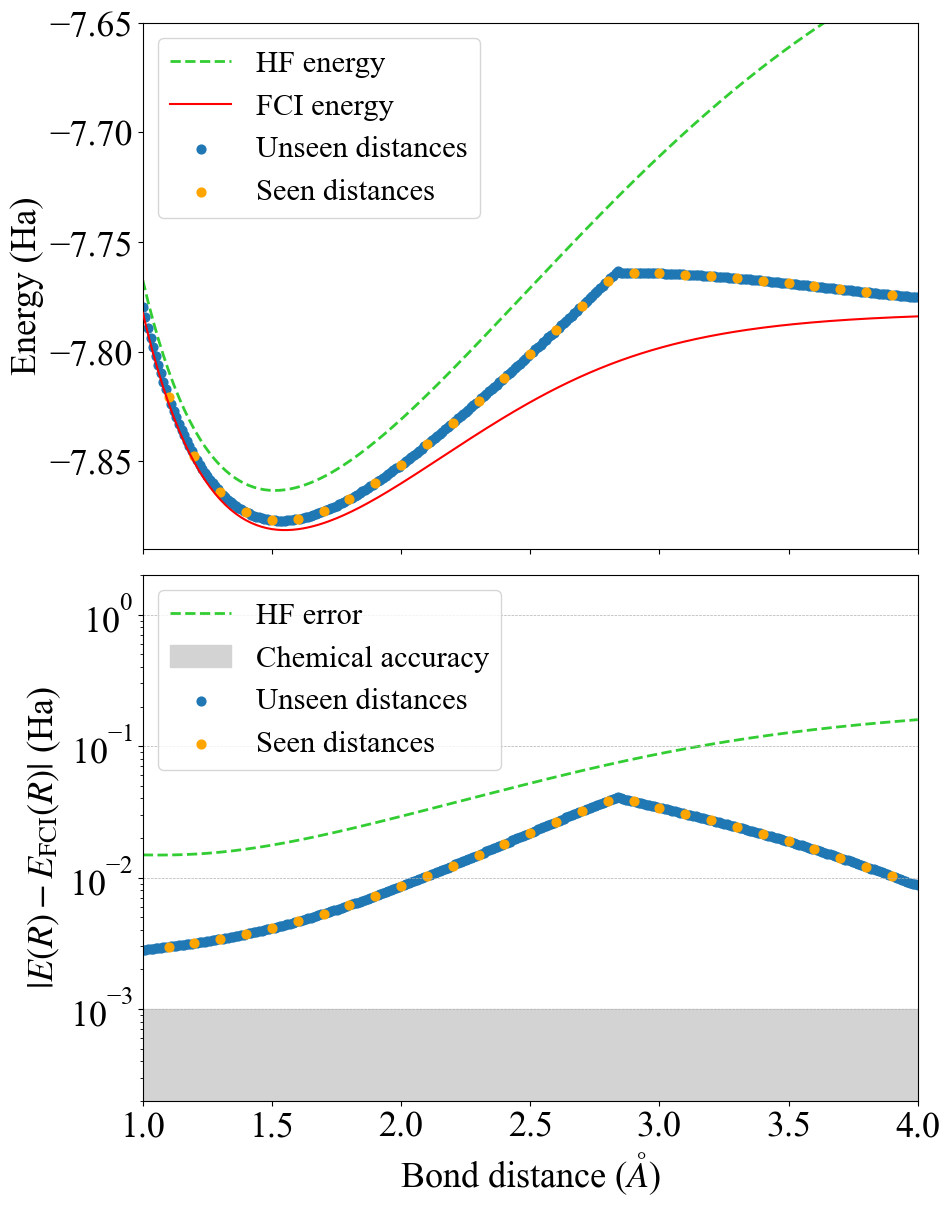}
    \caption{Potential energy curve for six-qubit LiH (top) and absolute energy differences from the FCI energy (bottom). The orange points represent the learned energies and errors at the bond distances used during training to condition the circuit on bond distance. The blue points correspond to the predicted energies and errors for unseen bond distances, reflecting the model's generalization. The red curve represents the FCI potential energy curve for the six-qubit LiH system, and the gray area indicates the chemical accuracy. The dashed green line represents the HF approximation of the PEC. }
    \label{LiH6_Transferability}
    \end{figure}
    
Fig.~\ref{LiH6_Transferability} shows one representative best-case PEC out of the twelve runs for six-qubit LiH (top) and the corresponding error relative to the FCI energy (bottom). This run achieves an average error of $0.0159^{+0.0132}_{-0.0092}$ Ha across the PEC, which is comparable to the accuracy achieved by the single-distance setup at $2.2$\,\AA\ (see Appendix~\ref{app:LiH6_supplementary_material}). By the same reasoning as in the four-qubit case, the predicted energies across the PEC can be associated with an UpCCD-level description, and the corresponding constructed circuits have depths in the range 9–12.

For LiH with a two-electron, six-orbital active space (CAS(2,6)) in Jordan-Wigner encoding, optimal circuits with a depth of 6 can represent the ground state exactly when the orbitals are chosen optimally (see Section~2.1 of Ref.~\cite{Kottmann_2023}). This orbital optimization was not applied to the Hamiltonian used here. If the orbital rotations are explicitly realized in the circuit, the depth increases to a range between 30 and 60, depending on the explicit encoding and level of optimization (here we used the compiling schemes of~\cite{santos_2025, yordanov2020efficient} as a stand-in for the state-of-the-art, but we can still imagine further optimizations). 
Gustiani \textit{et al.} reported a depth-14 circuit that prepares an approximation to chemical accuracy (see Fig.~5 in \cite{Gustiani_2023}), using a slightly larger system, LiH(4,12). On closer inspection, the circuit, however, only acts non-trivially on the LiH(2,6) subsystem used in this work. Taken together, this roughly confines the optimal circuit depth of our Hamiltonian to 6-14. Thus, although not exact, our 9–12-layer circuits lie within a depth-efficient window, particularly in the absence of orbital optimization.

In terms of training cost, we used 50,000 episodes in total, which effectively corresponds to approximately $1{,}724$ episodes per bond distance. This represents only a 1.24-fold increase in training cost per bond distance compared to training the PEC for four-qubit LiH, despite the discrete action space doubling and the state-space dimensionality increasing by a factor of four.

\subsection{Eight-qubit H\textsubscript{4}}
\label{PECH4}
Out of twelve runs, two runs generated a PEC for H$_4$ that achieved lower energies than the HF approximation across most bond distances in the range 0.5--1.55\,\AA. The details on all twelve runs can be found in Appendix~\ref{app:H4_supplementary_material_PEC}, and the corresponding hyper-parameters are specified in Appendix~\ref{app:hyper-parameters}. 
In Appendix~\ref{app:H4_supplementary_material_PEC}, we discuss the single-distance setup for H$_4$ at $1.5$\,\AA, where four out of twelve runs achieved a lower energy than HF, already indicating the difficulty of converging beyond 
the HF energy.

\begin{figure}[!htbp]
    \centering
    \includegraphics[width=0.95\columnwidth]{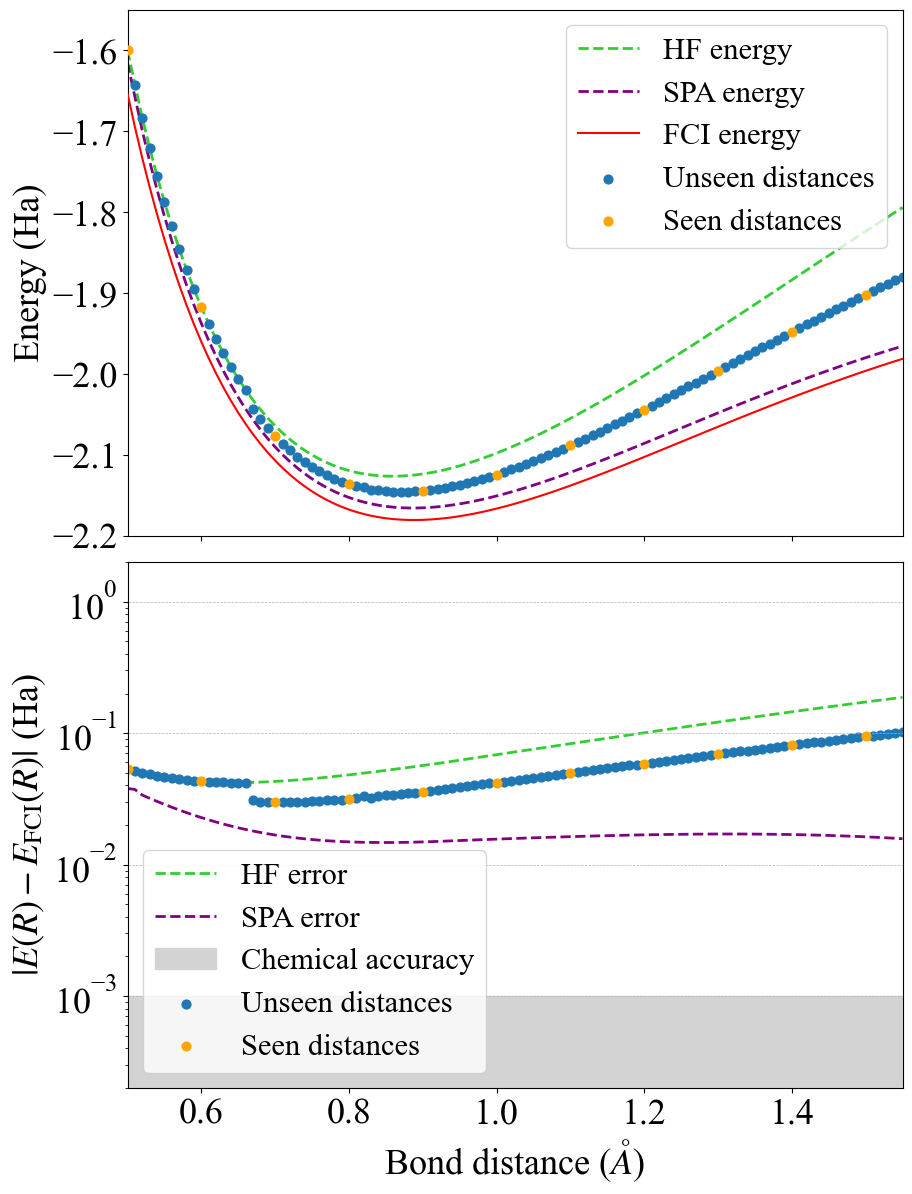}
    \caption{Potential energy curve for H$_4$ (top) and absolute energy differences from the FCI energy (bottom). The orange points represent the learned energies and errors at the bond distances used during training to condition the circuit on bond distance. The blue points correspond to the predicted energies and errors for unseen bond distances, reflecting the model's generalization. The red curve represents the exact potential energy curve for H$_4$, and the gray area marks chemical accuracy. The dashed green line represents the HF approximation of the PEC, while the dashed purple curve represents the SPA approximation of the PEC.
    }
    \label{H4_Transferability}
\end{figure}

Fig.~\ref{H4_Transferability} shows one representative best-case PEC out of the twelve runs for H$_4$ (top), along with the corresponding error relative to the FCI energy (bottom).
This run achieved an average error of ${0.0538}_{-0.0158}^{+0.0265}$ Ha across the PEC. On average, this is $1.7$ times more accurate than the HF approximation for the PEC (dashed green line), while remaining approximately $3.1$ times less accurate than the SPA approximation (dashed purple line).

The PEC for H$_4$ was trained on $11$ bond distances, using $50{,}000$ training episodes, which corresponds to $4{,}545$ episodes per bond distance. 
This again represents only a modest increase compared to the six-qubit LiH case, despite the discrete action space approximately doubling and the state space increasing by a factor of four.

\begin{figure}[ht!]
\includegraphics[width=0.95\columnwidth]{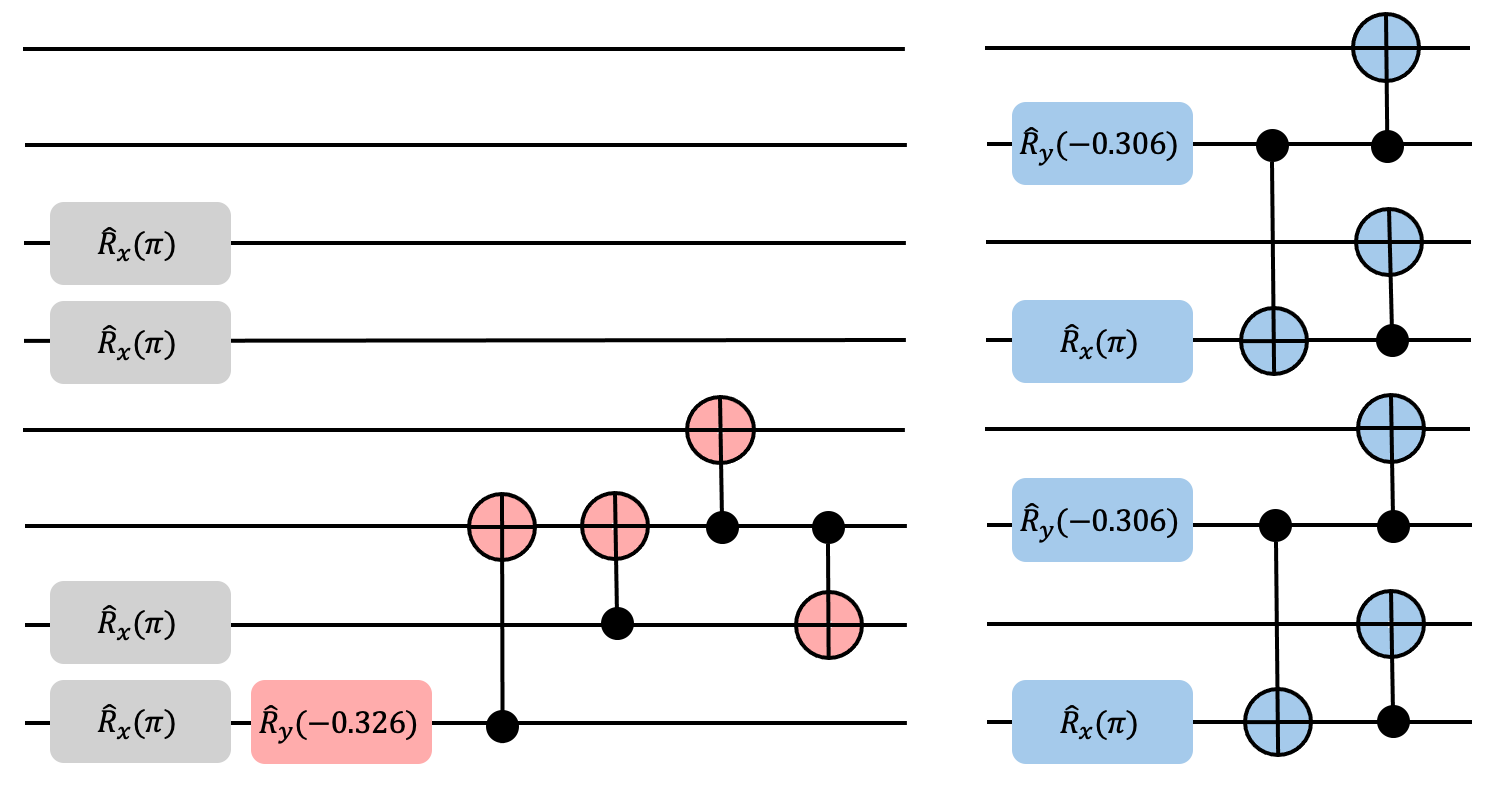}
\caption{Left: The generated circuit at a bond distance of 1.05\,\AA, shown in red, exemplifies the structure learned by the RL agent to entangle qubits four through seven across the entire bond distance range. Right: The SPA ansatz for H$_4$ at 1.05\,\AA, depicted in blue, composed of two separate entangling blocks.}
\label{circuit_H4}
\centering
\end{figure}

To analyze the circuits generated by the RL agents, they were further preprocessed as described in Appendix~\ref{circuits_pre-processing}. The circuits generated across the PEC do not resemble the SPA ansatz~\cite{Kottmann_2023}, which, despite being extremely shallow, was not identified by the stochastic sampling process of the RL agent, likely due to its highly symmetric structure. Nonetheless, an interesting entanglement pattern could be identified in the generated circuits, recurring at multiple bond distances: as exemplified by circuit Fig.~\ref{circuit_H4} for a bond distance of 1.05\,\AA, the agent consistently entangles only qubits four to seven, leaving the remaining four qubits untouched throughout. This behavior, while not reproducing the SPA ansatz explicitly, reflects a central idea underlying it: namely, the treatment of H$_4$ as two separate H$_2$ fragments, each forming an independent entangled block~\cite{Kottmann_2023}.  

Building on this observed structure, one could mask actions that couple the two blocks, thereby reducing the action space and potentially facilitating the agent's discovery of the full SPA ansatz.

\subsection{Qualitative insights} 

In this section, we share some observations and intuitions gathered throughout the development and evaluation of our framework that appeared to have the most pronounced impact on performance. 
These insights may offer helpful guidance to others working on related problems.

\begin{itemize}[topsep=0cm,left=0cm,itemsep=-0.8ex]
    \item  \textit{Qubit mapping:} The choice of qubit mapping affects the representation of the ground state, and we observed that this had a substantial impact on the performance of the RL agent. In particular, during early tests, we evaluated the Bravyi–Kitaev mapping with tapering for the four-qubit LiH, rather than the later-used parity mapping. Despite identical qubit counts, the RL agent converged more slowly and achieved noticeably worse final performance under the Bravyi–Kitaev mapping.  
    \item \textit{Starting state:} The starting state with the highest overlap to the exact solution was not always the best starting point for discovering lower-energy circuits, as we observed in our six-qubit LiH experiments. One possible explanation is that a "good" starting state may already lie in a well-defined local minimum, making it harder for the agent to explore alternative solutions.
    \item \textit{State representation:} The specific encoding of the quantum state as an input for the neural network critically influences the uniqueness of the state–action–reward mapping and, consequently, the overall complexity of the learning problem. In the literature, several implementations input the gate-level description directly \cite{ostaszewskiReinforcementLearningOptimization2021}. In contrast, we observed that circuit-based representations performed noticeably worse than the state representation. This performance gap can be explained by the ambiguity of circuit encodings relative to the state representation: different gate sequences can prepare the same quantum state, and a gate's effect depends strongly on its position within the circuit. The added complexity due to ambiguity seems crucial when the learning task is made more demanding by the additional challenge of optimizing continuous gate parameters and the objective of achieving generalization. However, this limitation might be mitigated by using more expressive classical models for circuit design, such as transformer architectures, which excel at capturing long-range dependencies in sequential data via self-attention mechanisms~\cite{li2023survey}.
    \item \textit{Reward design:} While reward design is recognized as a central aspect of RL, it is pivotal in our framework. The performance improved substantially when transitioning from a linear reward to an exponential reward, and further increased with a dynamically adapted reward. We attribute these improvements primarily to the adaptive sensitivity of the reward functions to the relevant energy scale, which allows the agent to distinguish small differences in energy based on gate choice and placement.  
    \item \textit{Gate set:} The difficulty of the learning task is determined by the action space, which in our case is defined by the gate set that we choose to be problem and hardware-agnostic. Initially, we conducted preliminary tests using single- and double-excitation gates on the six-qubit LiH system. These runs achieved convergence within chemical accuracy while requiring only about one-tenth as many training episodes as our HEA gate set. However, we maintained use of the HEA gate set for the sake of having a problem-agnostic and transferable framework, and to examine which structural patterns an RL agent can extract without prior knowledge of the system it is optimizing for. In general, however, altering the gate set seems to be the straightforward path to achieve performance increases. 
\end{itemize}

\section{Discussion and outlook} \label{discussion}
The performance of variational quantum algorithms critically depends on the choice of quantum circuit ansatz. Yet, no single ansatz can reliably generalize across different problem instances or even within a single parameter-dependent class, as ground-state properties can vary substantially with system parameters.
In this work, we introduce a reinforcement learning (RL) framework that formalizes transferable quantum circuit construction as a \emph{Markov decision process} (MDP) and implements it via the \emph{soft actor–critic} (SAC) algorithm. Given a molecular Hamiltonian class and a gate set, the RL agent is trained on a set of discrete bond distances. This enables the agent to generate individually adapted circuits for arbitrary bond distances within the continuous parameter range, thereby enabling direct access to the wavefunction and corresponding energy of the molecule under investigation. We evaluated our framework on the task of generating potential energy curves for LiH (in both four- and six-qubit representations) and H$_4$ (in an eight-qubit representation). Our results show consistent improvements over the \emph{Hartree–Fock} (HF) approximation, while substantially reducing the training cost per bond distance compared to independent single-point evaluations, as summarized in Table~\ref{tab:summary_results}.

The following discussion addresses the limitations, implications, and potential extensions of this approach.

\subsection{Caveats and drawbacks}

The deliberate choice of a problem-agnostic and domain-independent framework entails an inherent trade-off in performance. Thus, while not the primary goal of this work, our framework does not match the post-training accuracy of ansätze optimized for individual bond distances \cite{romero2018, ADAPT_VQE} or classical methods such as FCI or CCSD(T) \cite{jensen2007introduction}. The success of approaches of this kind generally relies on careful prior consideration of problem-specific symmetries to enable a suitable restriction of the Hilbert space~\cite{fedorov2021,cerezo2021VQAreview}. In our work, we deliberately adopted an entirely unconstrained setting by employing the \emph{hardware-efficient ansatz} HEA gate set. This way, we can precisely examine what circuit structures the agent can discover without any prior knowledge. This choice, however, substantially increases the difficulty of the learning problem, as it entails an (in principle) exponentially growing state space and a discrete action space whose size scales quadratically with the system size. Such conditions can hinder exploration and lead to weak reward–action correlations, which in turn may result in overfitting and other training pathologies~\cite{zhang2018,dulacarnold2016,Zhu_2022}. In future work, incorporating prior knowledge into the framework, such as chemically motivated gate sets, would reduce the complexity of the state and action spaces and is likely to yield significant improvements in standard performance metrics, such as accuracy.

A notable yet practical limitation is that our current framework is only feasible within classical simulation settings. Executing it on quantum hardware would necessitate full-state tomography at every step, resulting in an exponentially scaling measurement cost~\cite{Cramer_2010}. Adapting the framework for use on quantum hardware may be possible by changing the state and reward representations, for example, by encoding the circuit itself as the neural network input, as proposed in~\cite{ostaszewskiReinforcementLearningOptimization2021}. While this approach reduces the measurement requirements to a single evaluation per episode, it also introduces a sparse reward signal~\cite{riedmiller2018learning}. 
Additionally, it creates structural ambiguity, as different circuits can correspond to the same quantum state but are treated as distinct inputs by the neural network. Both factors introduce additional challenges for stable learning. 
Thus, while our method, in its current form, is restricted to classical simulation, similar ideas could be made scalable, for instance, by varying the approach or by extending it to more general settings, as explained below.

\subsection{Relevance and insights of the approach}

One of the main features demonstrated in our work is the inevitable shift in perspective from individual ansätze towards transferability by design. Specifically, the agent’s ability to identify and reuse shared structural patterns across different problem instances, while adapting them to the specifics of each case. This capability represents a distinctive advantage of learning-based approaches over traditional ansatz design methods. As we explicitly demonstrated, it can also significantly reduce the typical high measurement costs associated with ML techniques compared to single-point optimization. As an example, for four-qubit LiH in Section~\ref{PECLiH4}, we observed a 22-fold improvement in training efficiency for training the PEC compared to the single-bond-distance setup shown in Section~\ref{SBDLiH4}. We therefore argue that transferability should not be viewed merely as an optional benefit, but rather as a fundamental property that should be incorporated by default into \emph{machine learning} (ML)-based quantum circuit design methods. Moreover, we aim to highlight that generalization broadens the notion of what constitutes a useful solution. In many problem classes, obtaining the exact ground state can be highly challenging. In such cases, finding a generalizable solution may be more valuable, for example, as a warm start for further refinement~\cite{Puig_2025,foster2025abinitiofoundationmodel}. The knowledge gained from generalizable solutions about the molecule, compared to a specific ground state configuration, can accumulate additional system knowledge. 

A second important insight from our findings is that ML-based approaches need not be regarded solely as black-box optimizers; they inherently possess a degree of interpretability. While interpretation must be approached with caution, particularly in the presence of artifacts such as symmetry breaking, our results suggest that meaningful structural patterns can emerge. For example, our analysis of the generated circuits for the four-qubit LiH PEC revealed that $\hat{R}_x$ gates appear to play no meaningful role in constructing chemically relevant circuits, whereas $\hat{R}_y$ gates are essential and increasingly employed to account for stronger electron correlation. While such findings may seem trivial from the perspective of quantum chemistry, they reflect long-standing domain knowledge and were nevertheless recovered by the agent without hard-coded physical priors, within a reasonable training time of 40000 episodes. This highlights the potential of ML to yield interpretable, domain-consistent insights into circuit design. The potential interpretability may become particularly important in the context of scalability. As we discuss in the next section, scaling to larger systems may critically rely on understanding and reusing such structural patterns identified in smaller instances. 

We also want to highlight that numerically confirming the operation of the discrete-continuous SAC-based RL algorithm already yields a potential insight. In the literature, SAC is typically benchmarked either on purely continuous action spaces with dimensionalities of about 2–20 actions \cite{haarnoja2018,haarnoja2019} or on purely discrete settings with fewer than 20 actions \cite{christodoulou2019softactorcriticdiscreteaction}. By contrast, our approach combines both discrete and continuous actions, each comprising up to 24–80 actions, depending on the qubit number—a scale that exceeds standard benchmarks by a considerable margin \cite{Erdman_2023,delalleau2019}. This highlights the suitability of SAC for our task in terms of sample efficiency and motivates the exploration of clever SAC extensions and approaches. 
A central element in this is the replay buffer, which stores transitions to allow the agent to continuously learn from previously evaluated circuits without the need for repeated quantum measurements. This could not only further improve sample efficiency but also, in the future, support the parallelization of the data collection process. Transitions could, in principle, be collected across multiple quantum devices simultaneously, each constructing a circuit independently based on the current policy, and the resulting data could then be aggregated into a shared replay buffer from which the policy is updated.

Lastly, this work aims to highlight the conceptual value of RL as a fundamentally different paradigm for quantum circuit construction compared to the predominantly used gradient-based methods. The ability to explore circuit structures non-greedily, allowing intermediate increases in energy, has enabled the discovery of remarkably shallow circuit architectures, as demonstrated in Section~\ref{SBDLiH4} and also reported by other authors, such as in Ref.~\cite{ostaszewskiReinforcementLearningOptimization2021}. This conceptual difference is also an interesting aspect to investigate in the context of optimization landscape shaping, specifically the mitigation of barren plateaus \cite{mccleanBarrenPlateausQuantum2018b}. For ADAPT-VQE, the ability to sequentially reshape the optimization landscape at each step has been discussed in Ref.~\cite{Grimsley_2023} as a strategy to mitigate barren plateaus and rugged landscapes with local traps, which also applies to our setting. Unlike ADAPT-VQE, however, our approach does not enforce a monotonic energy reduction with every added gate.
The RL agent’s capacity to temporarily explore suboptimal circuit configurations, as illustrated in Fig.~\ref{LiH4_nongreedy}, can be particularly advantageous for escaping architectural traps and reshaping the energy landscape to access deeper minima. Investigating possible scenarios where this applies, however, lies beyond the scope of the present work.

\subsection{Aspects of scalability}
The scalability of quantum algorithms is especially critical in the variational setting, where the cost function requires repeated expectation-value estimation, with measurement costs scaling unfavorably with both system size and desired accuracy~\cite{cerezo2021VQAreview,2021arXiv211105176T}. This challenge is further exacerbated by the RL component, which is inherently costly. Compared to a standard VQE PEC interpolation scheme for four-qubit LiH (see, for example, the LiH PEC in Ref.~\cite{Kandala_2017}, where roughly 10 bond distances required about 250 iterations each), our RL framework required up to sixteen times more energy evaluations. This is especially critical, as estimating the cost scaling behavior with respect to system size and circuit depth is highly challenging. While this must always be put into perspective, given the rich insights RL offers, as discussed previously, refinements will be required to make our framework scalable. In the following, we will discuss promising ideas.

For instance, our framework could be applied to channel synthesis tasks, particularly to circuit compilation, where the cost function is not the energy in the variational problem itself, but rather the closeness to a given unitary (or more generally, a target channel) as input. In such settings, \emph{sliding window} optimization can be employed: only a fixed number of the most recently added gates are optimized at a time, while earlier gates remain fixed. This maintains the optimization cost per step as constant, independent of the total circuit depth, thereby ensuring scalability. Recent work in the circuit compilation setting has already demonstrated that RL agents can generalize circuit components learned on smaller systems to larger qubit counts~\cite {nakaji2025}.  We hope that our current work, with its technical developments, will stimulate further steps that our framework can, in principle, generate generalizable circuits for a wide range of objectives by simply adapting the reward function.

It is also worth mentioning that while the potential of near-term quantum computing approaches~\cite{bharti2021noisy,preskillQuantumComputingNISQ2018a} has been viewed as overly optimistic in recent years, the community currently tends to be overly pessimistic: the intermediate regime preceding the advent of fault-tolerant quantum computers should be fully assessed. Besides, variational quantum algorithms are expected to play a role even in the fault-tolerant era \cite{Myths,MindTheGaps}. For instance, a promising direction for 
 improving scalability in the variational setting involves fragmentation methods, which could also be integrated into our framework. These methods aim to reduce circuit complexity by decomposing the problem into smaller, tractable subsystems. 
 Recent work~\cite{farrell2024,Schleich_2023} has demonstrated that scalable ansätze can be constructed by assembling circuits from such fragments. Specifically, our framework can be applied to individual fragments to identify non-intuitive circuit structures through non-greedy optimization. These could either be assembled manually into larger ansätze or reintroduced into the action space as fixed building blocks when scaling to larger systems. Further, the replay buffer could also serve as a repository of the structural information shared between building blocks, thereby enabling information-based scaling. Such modular reuse may provide a practical and efficient pathway toward constructing circuits for larger quantum systems within our framework.

\subsection{Outlook}
Our findings highlight RL as a promising and versatile tool for ansatz design, particularly by enabling transferability, revealing structural patterns, and leveraging its inherently non-greedy exploration strategy. As discussed above, achieving \emph{full scalability} is key to advancing RL-based circuit design. 
Several promising directions discussed here may support this goal in future work. In particular, we hope that our technical implementation facilitates adaptation to other problem domains beyond quantum chemistry and entirely different tasks such as channel synthesis. Moreover, we particularly encourage the development of methods that further exploit and systematically reuse the agent's building blocks and learned knowledge, for example, as discussed in Section~\ref{PECLiH4}, to enable scaling to larger systems. This could be realized through mechanisms such as sharing knowledge via the replay buffer, reintroducing learned structures into the action space, or assembling them through fragmentation-based methods. Lastly, we see particular potential to establish non-greedy RL circuit discovery—as characteristically illustrated by the plot in Fig.~\ref{LiH4_nongreedy}—through a deeper theoretical investigation of the advantages of non-greedy exploration. This could be pursued, for example, by analyzing concrete scenarios in which RL outperforms greedy methods by traversing suboptimal parameter regions, or by examining how non-greedy behavior can aid in a more accessible interpretation of the resulting circuits.

\section*{Data Availability}
The code used to obtain our numerical results is available on GitHub at~\href{https://github.com/porscheofficial/reinforcement-learning-of-quantum-circuit-architectures/tree/main}{porscheofficial/reinforcement-learning-of-quantum-circuit-architectures}, and the corresponding data is available via Zenodo at \href{https://doi.org/10.5281/zenodo.17567243}{DOI: 10.5281/zenodo.17567243}. All quantum circuits were simulated using Qulacs~\cite{Suzuki_2021}. PyTorch~\cite{paszke2019pytorch} with GPU acceleration was used for neural network training.

\textit{The results, opinions and conclusions expressed in this publication are not necessarily those of Porsche Digital GmbH.}

\begin{acknowledgments}
We acknowledge the computing resources provided by \href{https://aws.amazon.com}{AWS computing services} and the HPC Service of FUB-IT, Freie Universität Berlin, for this research. We thank Regina Kirschner, Mahdi Manesh, Peter Wolf, Stefan Zerweck, and Mattias Ulbrich, and Sofiene Jerbi for valuable discussions. We also thank Ed Barnes, Sophia Economou, Nick Mayhall, Frank No\'e, Zeno Sch\"atzle, and P. Bern\'at Szab\'o for fruitful discussions and insights. We thank the Federal Ministry of Research, Technology and Space (BMFTR) (HYBRID++, QuSol, MUNIQC-Atoms), as well as the Munich Quantum Valley, the QuantERA (HQCC), Berlin Quantum, and the European Research Council, for financial support. MK and AW gratefully acknowledge Porsche Digital GmbH for their support. PAE gratefully acknowledges funding by the Berlin Mathematics Center
MATH+ (AA2-18). SK acknowledges financial support from the US Department of Energy, Office of Science, Advanced Scientific Computing Research program, under award number DE-SC0025430.

\end{acknowledgments}

\newpage

\bibliography{main}

\newpage

\appendix

\onecolumngrid

\section{Supplementary material on setups}
\label{app:statistics}

This section provides additional information on the statistical evaluation methods and detailed results for each of the molecular systems considered in the main text: four-qubit LiH, six-qubit LiH, and eight-qubit H$_4$.

\subsection{Evaluation methods}
\label{app:statistical_methods}

\subsubsection{Statistical evaluation} \label{stats}
The mean error $\bar{x}_k$ of a single PEC run $k$ is computed as
\begin{align}
    \bar{x}_k = \frac{1}{M} \sum_{i=1}^{M} \left| E(R_i)-E_{\text{FCI}}(R_i) \right|,
\end{align}
where $E(R_i)$ denotes the energy predicted by the RL agent at bond distance $R_i$, $E_{\text{FCI}}(R_i)$ is the FCI energy at bond distance $R_i$, and $\{R_i\}_{i=1}^{M}$ is a discrete set of $M$ bond distances uniformly sampled over the interval $[R_{\text{min}}, R_{\text{max}}]$.

The mean PEC error $\bar{x}_{\text{runs}}$ across $N$ independent runs is then given by
\begin{align}
    \bar{x}_{\text{runs}} = \frac{1}{N} \sum_{k=1}^{N} \bar{x}_k.
\end{align}

To report performance variability across runs, we use the asymmetric standard deviation $\sigma_{-,+}$, which is defined for a set of values $\{x_i\}_{i=1}^N$ with mean $\bar{x}$ as
\begin{align}
    \sigma_{-} &= \sqrt{\frac{1}{N_{-} - 1} \sum_{i: x_i < \bar{x}} (\bar{x} - x_i)^2}, \\
    \sigma_{+} &= \sqrt{\frac{1}{N_{+} - 1} \sum_{i: x_i > \bar{x}} (x_i - \bar{x})^2},
\end{align}
where $N_{-}$ and $N_{+}$ are the number of samples below and above the mean $\bar{x}$, respectively.

The asymmetric standard deviation prevents violations of the variational principle, as large upward outliers can increase the standard deviation, potentially causing a symmetric error interval to fall below the true ground-state energy.

\subsubsection{Pre-processing of constructed circuits}
\label{circuits_pre-processing}
The agent must construct circuits with a fixed number of gates and cannot terminate circuit construction early. As a result, it may use part of its action budget to place redundant gate sequences, thereby indirectly indicating that fewer gates would suffice or that additional gates would not improve the outcome. To simplify the interpretation of the resulting circuits, we apply the following pre-processing steps to the agent's constructed circuits (when indicated in the text):
\begin{itemize}
    \item The sequence CNOT$(i,j)$CNOT$(l,k)$ is removed if $i=l$ and $j=k$.
    \item CNOT$(i,j)$ is removed if the control qubit is in state $\ket{0}$.
    \item Sequences of identical rotation gates acting consecutively on the same qubit are merged, e.g., $\hat{R}_x(\theta_1)\hat{R}_x(\theta_2)=\hat{R}_x(\theta_1+\theta_2)$.
\end{itemize}

\subsection{Four-qubit LiH}
The following section provides additional information and statistical evaluations for the single-bond-distance simulation at 2.2\,\AA\, and the potential-energy-curve (PEC) simulation over the bond-distance range $1.0$ to $4.0$\,\AA\, for the four-qubit LiH system. Further details on the molecular system can be found in Appendix~\ref{app:quantumchem}, and the hyper-parameters used for both the single-distance and PEC simulations are listed in Table~\ref{app:hyper-parameters}.

\subsubsection{Four-qubit LiH at 2.2\,\AA}
\label{app:LiH4_supplementary_material}

\begin{figure}[htbp]
    \centering
    \includegraphics[width=0.9\textwidth]{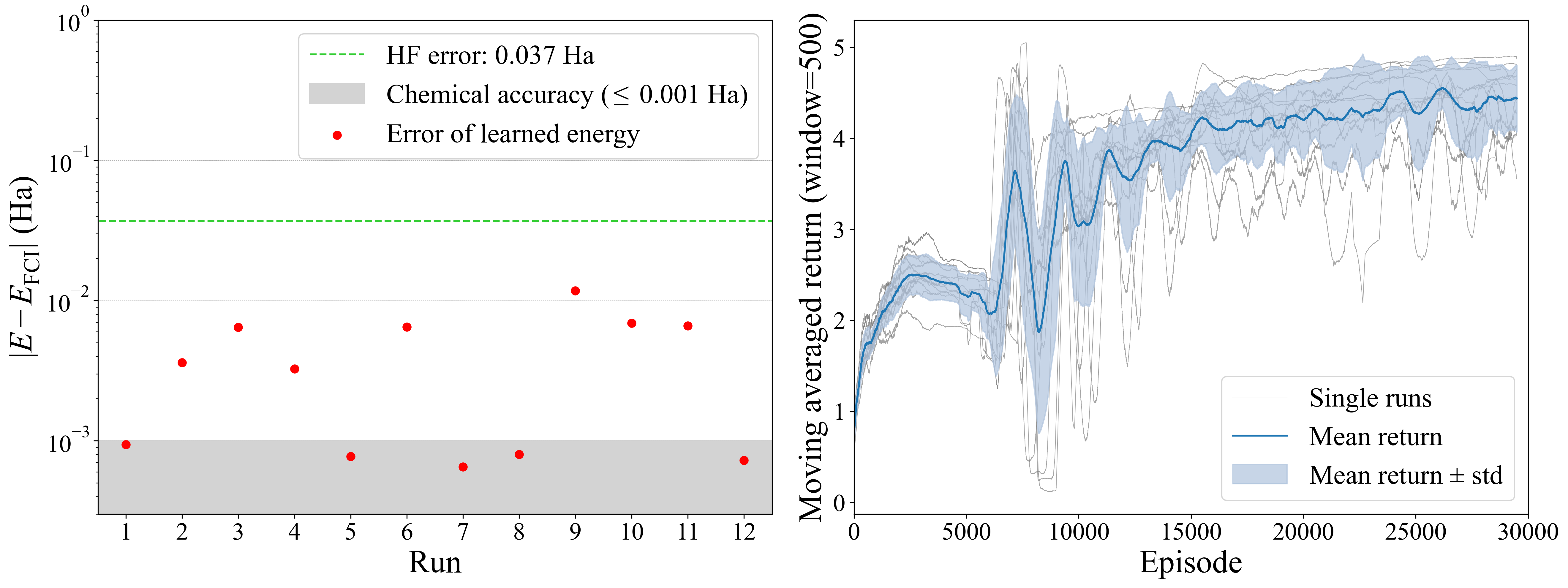}    
    \caption{Left: Learned energies for all twelve runs for the four-qubit LiH. Right: Moving averaged return with window size $500$ for all twelve runs for the four-qubit LiH.}
    \label{fig:energies_and_returns_LiH4}
\end{figure}

For the four-qubit LiH system at a bond distance of 2.2\,\AA, we conducted twelve independent runs using the fixed hyper-parameters specified in Table~\ref{tab:hyperparams}. Circuit construction was initiated from the HF state, which yielded better results than both the vacuum and random initial states. 
The mean learned energy across all runs was 
$-7.8408^{+0.0046}_{-0.0030}$ Ha, deviating from the FCI energy by 0.0041 Ha. The left panel of Fig.~\ref{fig:energies_and_returns_LiH4} shows the learned energy from each of the twelve simulated runs. 
The right panel of Fig.~\ref{fig:energies_and_returns_LiH4} displays the moving averaged return with window size $500$ $\pm$ one standard deviation across the twelve runs.

The increasing return indicates that the agent is still in the learning phase, while the plateau towards the end suggests convergence and that learning has effectively saturated.
The recurring oscillations in the return result from dynamic reward adaptation, which can lead to temporary drops in return before the neural network parameters readjust to the updated reward evaluation. A best-case result from the twelve runs was presented in the main text in Section~\ref{SBDLiH4}.

\subsubsection{Four-qubit LiH from 1.0\,\AA\, to 4.0\,\AA}
For four-qubit LiH in the bond distance range 1--4\,\AA, we performed twelve independent runs using fixed hyper-parameters, listed in Appendix~\ref{app:hyper-parameters}. The hyper-parameter choices were guided by those previously shown to work well in the single bond distance setting~\ref{app:LiH4_supplementary_material}, such as initializing from the HF state and using a maximal gate count of 12 gates, while slightly adjusting others, like the number of training episodes and the replay buffer capacity, to better accommodate the PEC setup. The left panel of Fig.~\ref{fig:PEC12_LiH4} shows the mean PEC along with one asymmetric standard deviation (see Appendix~\ref{stats}) across the bond distance range 1--4\,\AA, while the right panel of Fig.~\ref{fig:PEC12_LiH4} displays the corresponding mean error relative to the FCI energy on a logarithmic scale across the twelve runs. The orange energies and errors correspond to bond distances used during training, while the blue points indicate unseen bond distances, for which energy and error were predicted via generalization.
The PEC exhibits a mean error of $0.0136^{+0.0156}_{-0.0096}$ Ha, representing an improvement by a factor of approximately 5.1 over the HF approximation. The asymmetrical standard deviation is based on a limited number of samples. It can therefore exceed the HF error margin in the bond distance range 1--1.9\,\AA, even though none of the twelve runs actually performed worse than HF within this range (see Appendix~\ref{stats}). Around the point of symmetry breaking, the variance increases substantially due to the transition occurring at slightly different bond distances in each run. This occasionally leads to outliers that affect the mean and are responsible for the observed asymmetry in the standard deviation.

\begin{figure}[htbp]
    \centering
    \includegraphics[width=\textwidth]{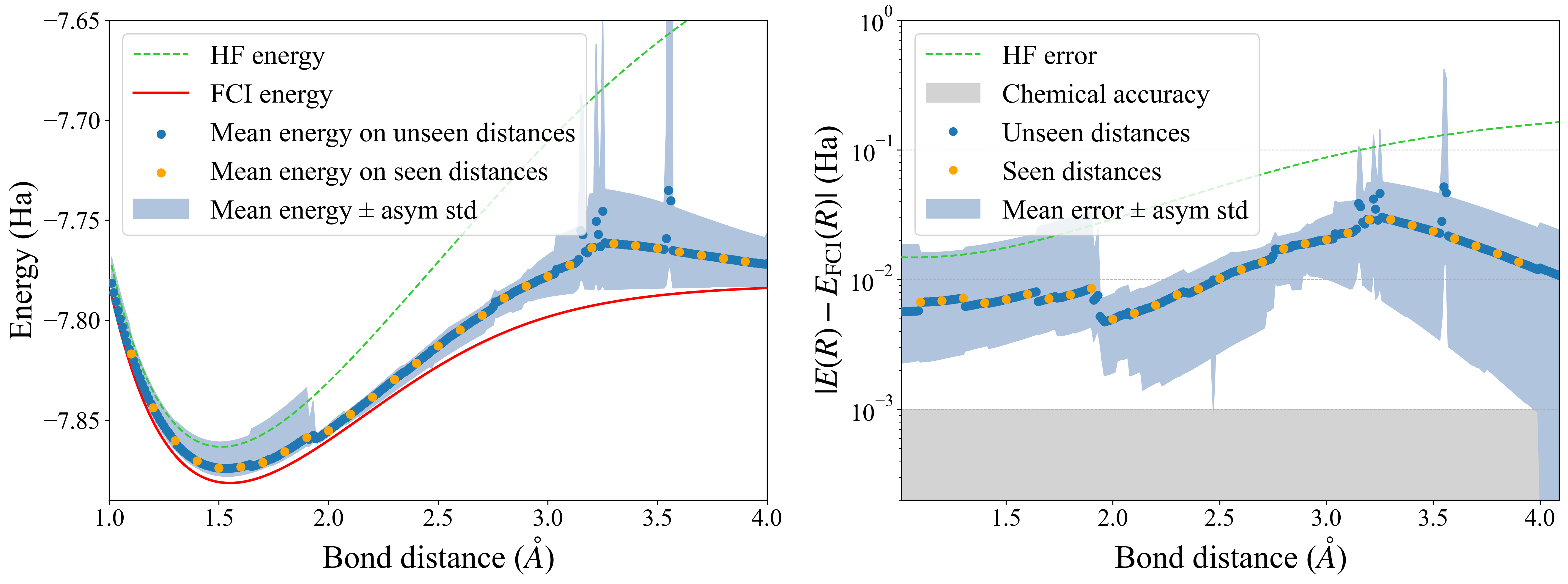}    
     \caption{Left: Mean predicted PEC across the twelve runs, with shaded region indicating one asymmetric standard deviation across the bond distance range 1--4\,\AA. Right: Mean energy error relative to the FCI energy, shown on a logarithmic scale with the shaded region representing one asymmetric standard deviation.}
    \label{fig:PEC12_LiH4}
\end{figure}

\begin{figure}
    \centering
    \includegraphics[width=0.5\linewidth]{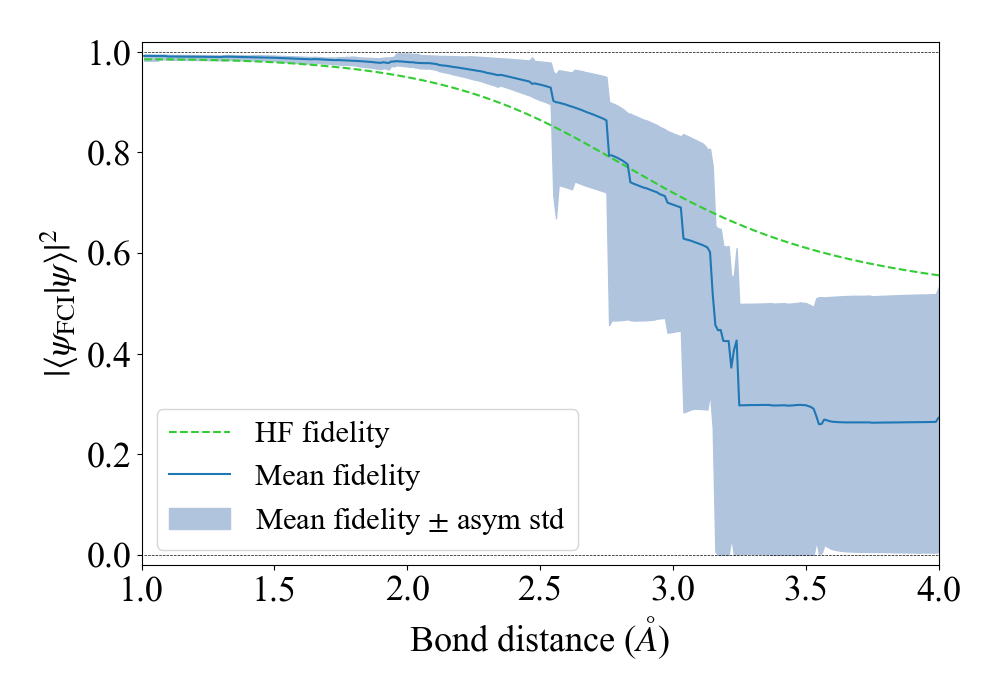}
    \caption{Mean fidelity of the predicted state with respect to the FCI ground state across all twelve runs, shown over the bond distance range 1--4\,\AA. The shaded area indicates one asymmetric standard deviation. The dashed green line shows the fidelity of the HF state with respect to the FCI ground state.}
    \label{fig:LiH4T_mean_fidelity}
\end{figure}

\paragraph{Mean fidelity of the predicted state with respect to the ground state across the PEC.}
Fig.~\ref{fig:LiH4T_mean_fidelity} shows the mean fidelity between the predicted state and the FCI ground state across all twelve runs over the bond distance range 1--4\,\AA. Throughout the range from approximately $1.0$\,\AA \,to $2.8$\,\AA, the predicted fidelity consistently exceeds that of the HF state (green dashed line). However, in each of the twelve runs, a sharp drop in fidelity occurs at some point within the interval between approximately $2.8$\,\AA\, and $3.2$\,\AA. As illustrated by the PEC in Fig.~\ref{fig:PEC12_LiH4}, this drop is not accompanied by a comparable deterioration in energy accuracy, which remains significantly better than the HF energy across the entire PEC. This indicates that, in this region, the agent transitions from a physically meaningful, symmetry-preserving state to a symmetry-broken solution that yields a low energy but poorly overlaps with the true ground state. 

\begin{figure}
    \centering
    \includegraphics[width=0.5\linewidth]{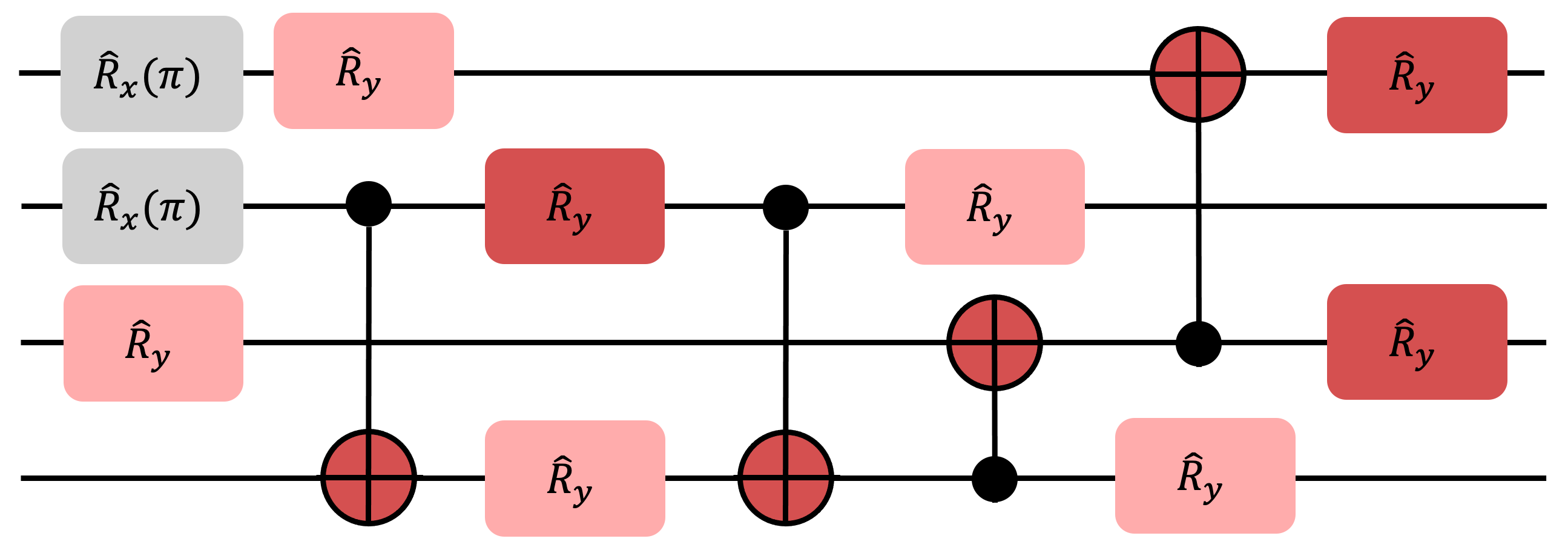}
    \caption{Schematic illustration of how the circuit architecture evolves within the bond distance range from $1.0$\,\AA\, to $2.8$\,\AA\, for the representative best-case run shown in Figure~\ref{LiH4_Transferability}. The gray $\hat{R}_x$ gates initialize the system in the HF state. Dark red gates represent the core circuit structure learned at $1.0$\,\AA, while light red $\hat{R}_y$ gates are gradually added with increasing bond distance. Circuit parameters are not visualized, but smoothly adapt at each bond distance. Beyond $2.5$\,\AA, the initial CNOT gate is removed. After $2.8$\,\AA, the architecture changes significantly due to symmetry breaking.}
    \label{fig:LiH4T_circuit_scheme}
\end{figure}

\paragraph{Schematic illustration of the circuit architecture.}
Fig.~\ref{fig:LiH4T_circuit_scheme} schematically illustrates how the circuit architecture evolves within the bond distance range from $1.0$\,\AA\, to $2.8$\,\AA\, for the representative best-case run shown in Fig.~\ref{LiH4_Transferability}.

The gray $\hat{R}_x$ gates initialize the system in the HF state and are applied identically across all bond distances. The dark red gates denote the core circuit structure learned at a bond distance of $1.0$\,\AA, which is largely preserved up to $2.8$\,\AA. While circuit parameters are not visualized, they are smoothly adapted at each bond distance. Light red $\hat{R}_y$ gates are progressively added between $1.0$\,\AA\, and $2.3$\,\AA. Beyond $2.5$\,\AA, the agent reduces the core entanglement structure by removing the initial CNOT gate. After the onset of symmetry breaking at 2.8\,\AA, this architecture is replaced by a significantly simplified circuit.

\subsection{Six-qubit LiH}
The following section provides additional information and statistical evaluations for the single-bond-distance simulation at 2.2\,\AA\, and the PEC simulation over the bond-distance range $1.0$ to $4.0$\,\AA\, for the six-qubit LiH system. Further details on the molecular system can be found in Appendix~\ref{app:quantumchem}, and the hyper-parameters used for both the single-distance and PEC simulations are listed in Table~\ref{app:hyper-parameters}.

\subsubsection{Six-qubit LiH at 2.2\,\AA}
\label{app:LiH6_supplementary_material}

\begin{figure}[t!]
    \centering
    \includegraphics[width=0.9\textwidth]{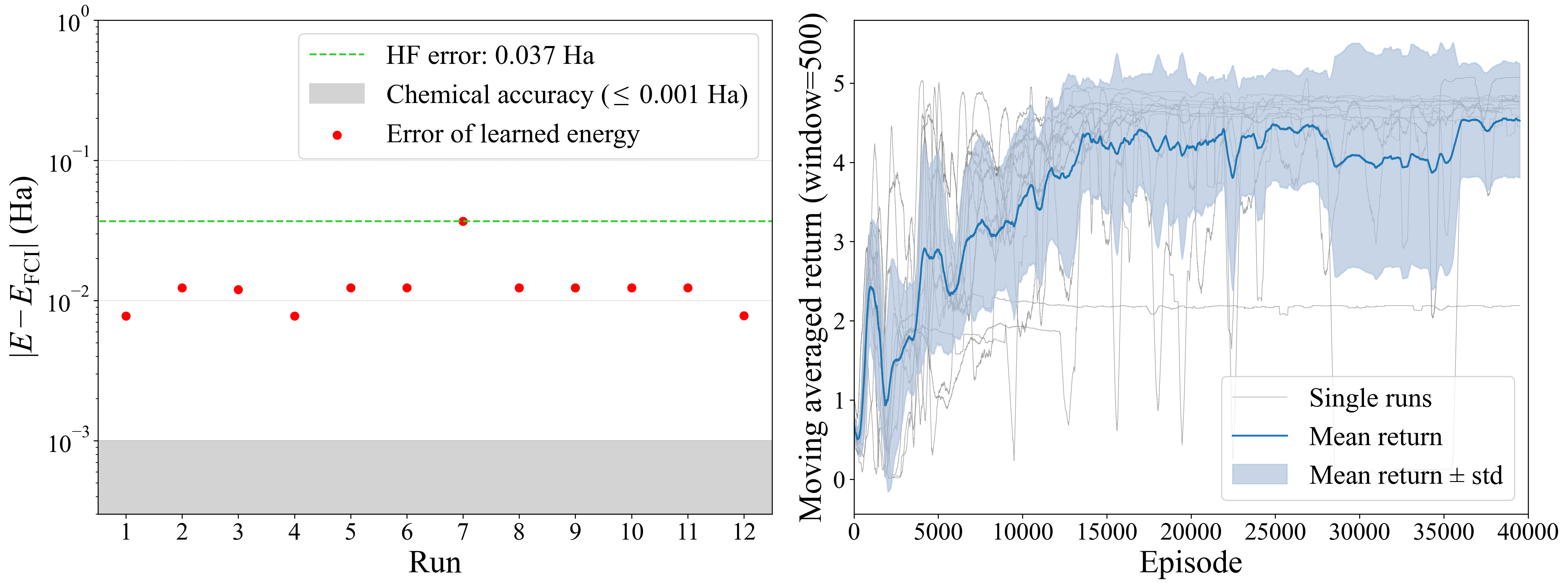}    
    \caption{Left: Learned energies for all twelve runs for the six-qubit LiH. Right: Moving averaged return with window size $500$ for all twelve runs for the six-qubit LiH.}
    \label{fig:energies_and_returns_LiH6}
\end{figure}

\begin{figure}[htbp]
    \centering
    \includegraphics[width=\textwidth]{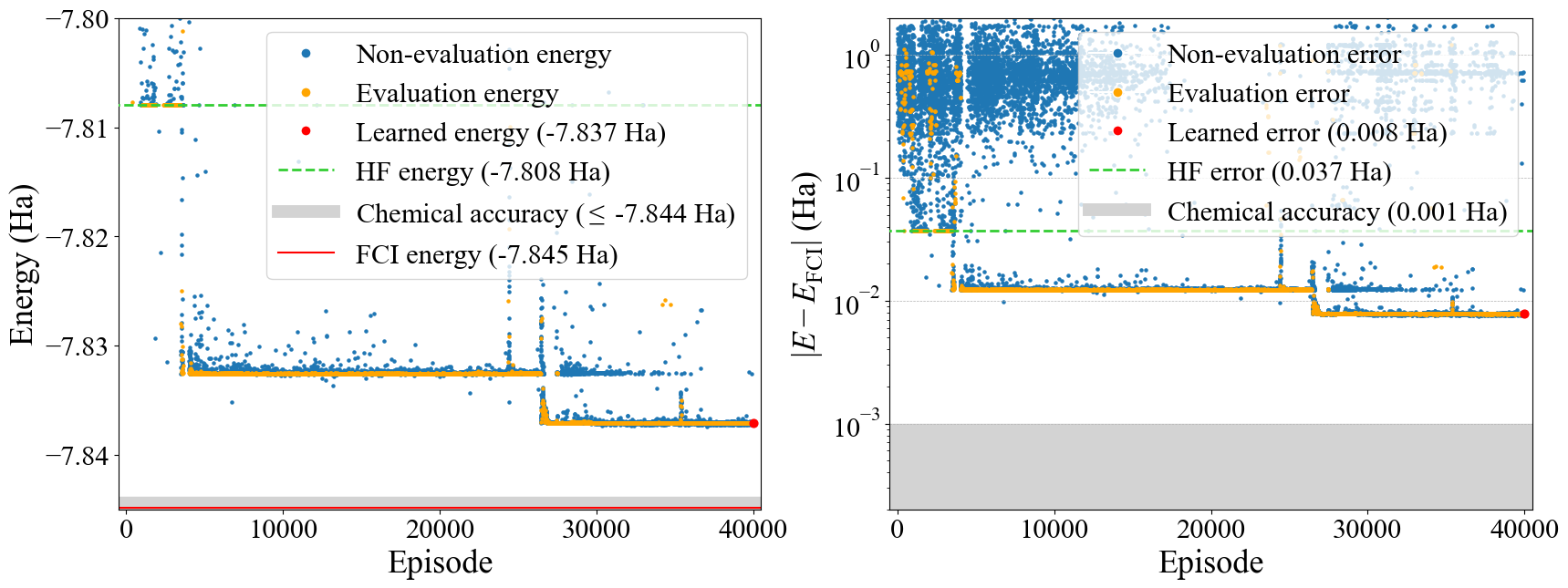}    
    \caption{Left: Progression of the obtained energies during training for the best-case run of the six-qubit LiH system.
    Right: Energy error relative to the FCI energy during training, shown on a logarithmic scale.}
    \label{fig:best_case_LiH6}
\end{figure}

For the six-qubit LiH system at a bond distance of 2.2\,\AA, we conducted twelve independent runs using the fixed hyper-parameters specified in Appendix~\ref{app:hyper-parameters}. Circuit construction was initiated from an arbitrarily chosen starting state (see Appendix~\ref{app:hyper-parameters}), as initializing from the HF state caused the agent to become stuck in that configuration. This behavior is likely due to the HF state representing a strong local minimum, which, although energetically favorable, limits the exploration of alternative solutions. In contrast, starting from a less energetically favorable state helped the agent escape local traps and ultimately led to lower energy states. The left panel of Fig.~\ref{fig:energies_and_returns_LiH6} displays the learned energies from all twelve independent runs for six-qubit LiH at a bond distance of 2.2\,\AA. The mean learned energy is $-7.8317_{-0.0031}^{+0.0237}$ Ha, corresponding to an average error of $0.0110$ Ha relative to the FCI energy. The right panel of Fig.~\ref{fig:energies_and_returns_LiH6} displays the moving averaged return with window size $500$ $\pm$ one standard deviation across the twelve runs. In line with all other setups, Fig.~\ref{fig:best_case_LiH6} shows one representative best-case run for the six-qubit LiH.  The final learned energy for this run is $-7.8371$ Ha, corresponding to an error of $0.0078$ Ha relative to the FCI energy.

\subsubsection{Six-qubit LiH from 1.0\,\AA\, to 4.0\,\AA}
\label{app:LiH6T}
\begin{figure}[htbp]
    \centering
    \includegraphics[width=\textwidth]{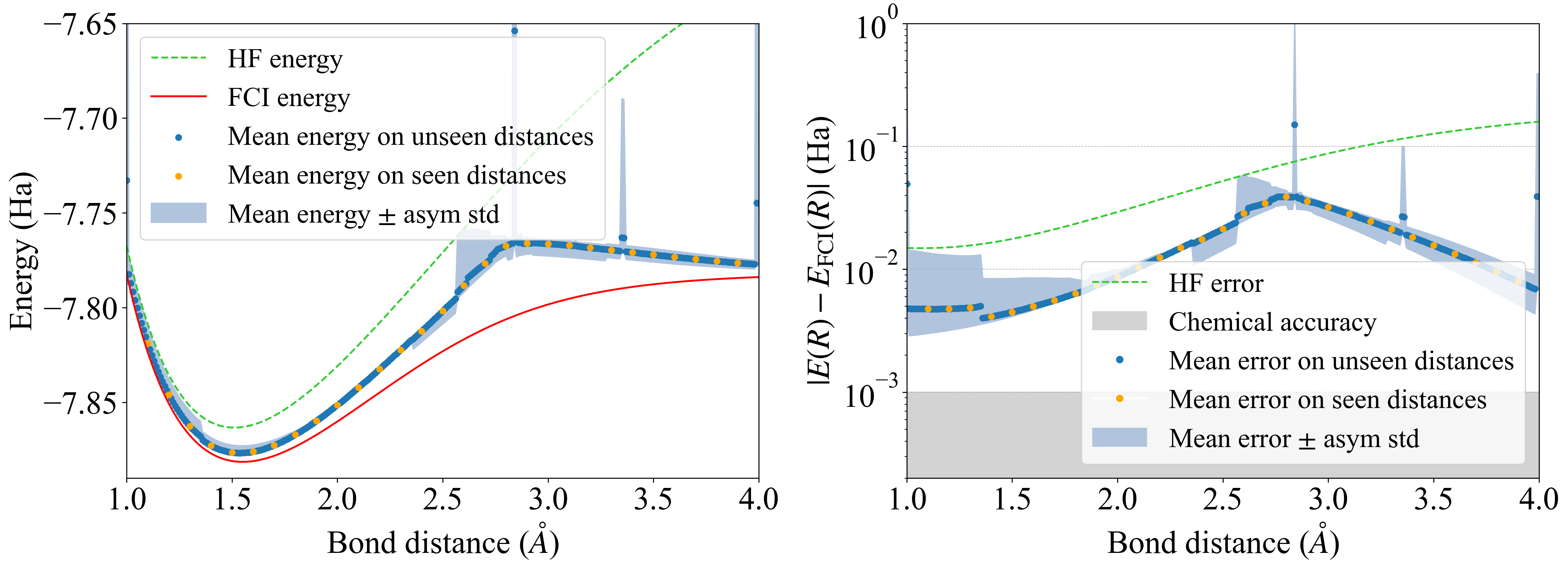}   
     \caption{Averaged PEC (left) and averaged error with respect to the FCI energy (right) across twelve runs for the six-qubit LiH. The shaded region represents the asymmetric standard deviation around the mean. The orange points represent the averaged learned energies and errors at the bond distances used during training to condition the circuit on bond distance. The blue points correspond to the averaged predicted energies and errors obtained for unseen bond distances, generalization of the model.}
    \label{fig:PEC12_LiH6}
\end{figure}

For six-qubit LiH in the bond distance range 1--4~\AA, we performed twelve independent runs using fixed hyper-parameters, listed in Appendix~\ref{app:hyper-parameters}. The hyper-parameter choices were guided by those previously shown to work well in the single bond distance setting (see Appendix~\ref{app:LiH6_supplementary_material}), such as initializing from an arbitrarily chosen state, while slightly adjusting others, like the number of training episodes and the replay buffer capacity, to better accommodate the transferable setup. The left panel of Fig.~\ref{fig:PEC12_LiH6} shows the mean PEC with one asymmetric standard deviation, while the right panel of Fig.~\ref{fig:PEC12_LiH6} displays the corresponding mean error relative to the FCI energy on a logarithmic scale. The orange energies and errors correspond to bond distances used during training, while the blue points indicate unseen bond distances, for which energy and error were predicted via generalization.
The average achieved accuracy across the PEC is $0.0161^{+0.0136}_{-0.0036}$ Ha, which is approximately 4.3 times better than the HF approximation. 
Outliers at $1.0$\,\AA\, and $4.0$\,\AA\, arise because they are farther from the nearest training points, with no further data beyond them to guide the model. Another outlier is observed near the symmetry-breaking region, where the abrupt transition in the ground state makes it more difficult for the model to produce smooth predictions.

\subsection{H\textsubscript{4}}
\label{app:H4_supplementary_material}
The following section provides additional information and statistical evaluations for the single-bond-distance simulation at 1.5\,\AA\, and the potential-energy-curve (PEC) simulation over the bond-distance range $0.5$ to $1.6$\,\AA\, for H$_4$. Further details on the molecular system can be found in Appendix~\ref{app:quantumchem}. The hyper-parameters used for both the single-distance and PEC simulations are listed in Table~\ref{tab:hyperparams}.

\subsubsection{H\textsubscript{4} at 1.5\,\AA} \label{app:H4_supplementary_material_single}

\begin{figure}[htbp]
    \centering
        \centering
        \includegraphics[width=0.9\textwidth]{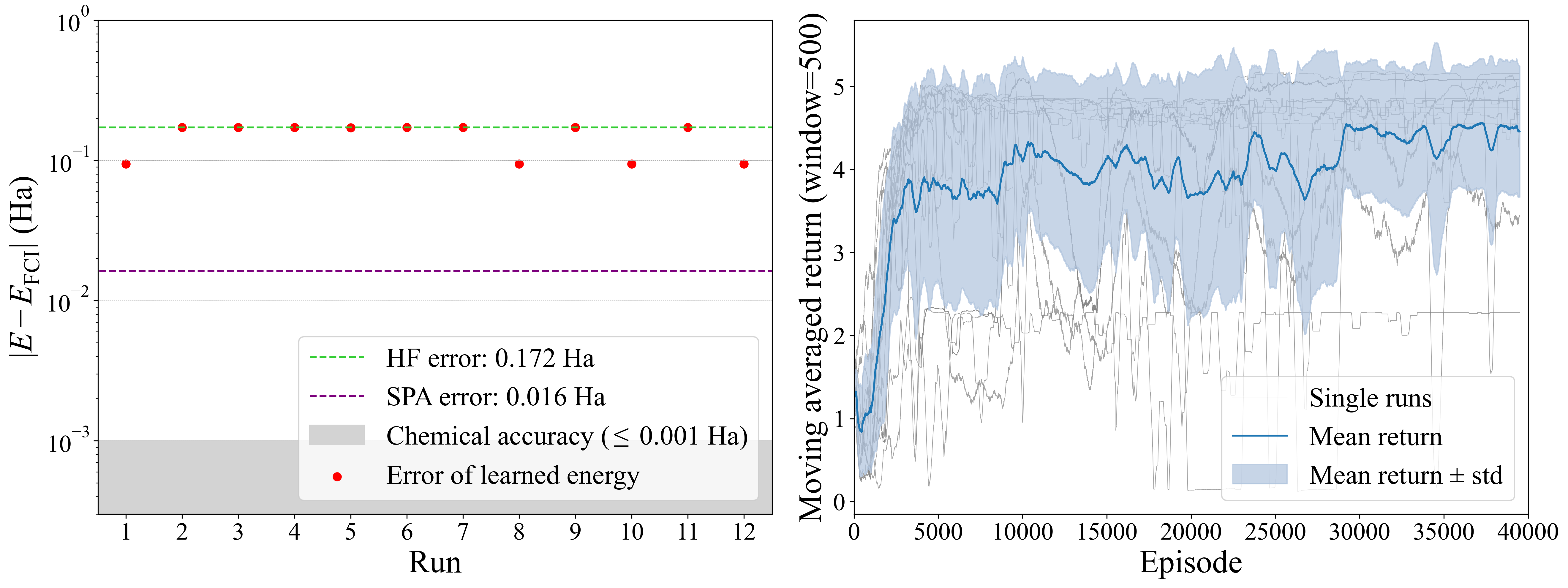}
    \caption{Left: Learned energies for all twelve runs for H$_4$ at a bond distance of 1.5\,\AA. Right: Moving averaged return with window size $500$ for all twelve runs for H$_4$ at a bond distance of 1.5\,\AA.}
    \label{fig:energies_and_returns_H4}
\end{figure}

The left panel of Fig.~\ref{fig:energies_and_returns_H4} shows the learned energy for all twelve simulated runs for H$_4$ at a bond distance of 1.5\,\AA\, under the specified hyper-parameters listed in Appendix~\ref{app:hyper-parameters}. The mean learned energy across these twelve runs is $-1.8499_{-0.0600}^{+0.0278}$ Ha, corresponding to a mean error of $0.1463$ Ha.

The right panel of Fig.~\ref{fig:energies_and_returns_H4} displays the moving averaged return with window size $500$ $\pm$ one standard deviation across the twelve runs. 

\begin{figure}[htbp]
        \centering
        \includegraphics[width=0.9\textwidth]{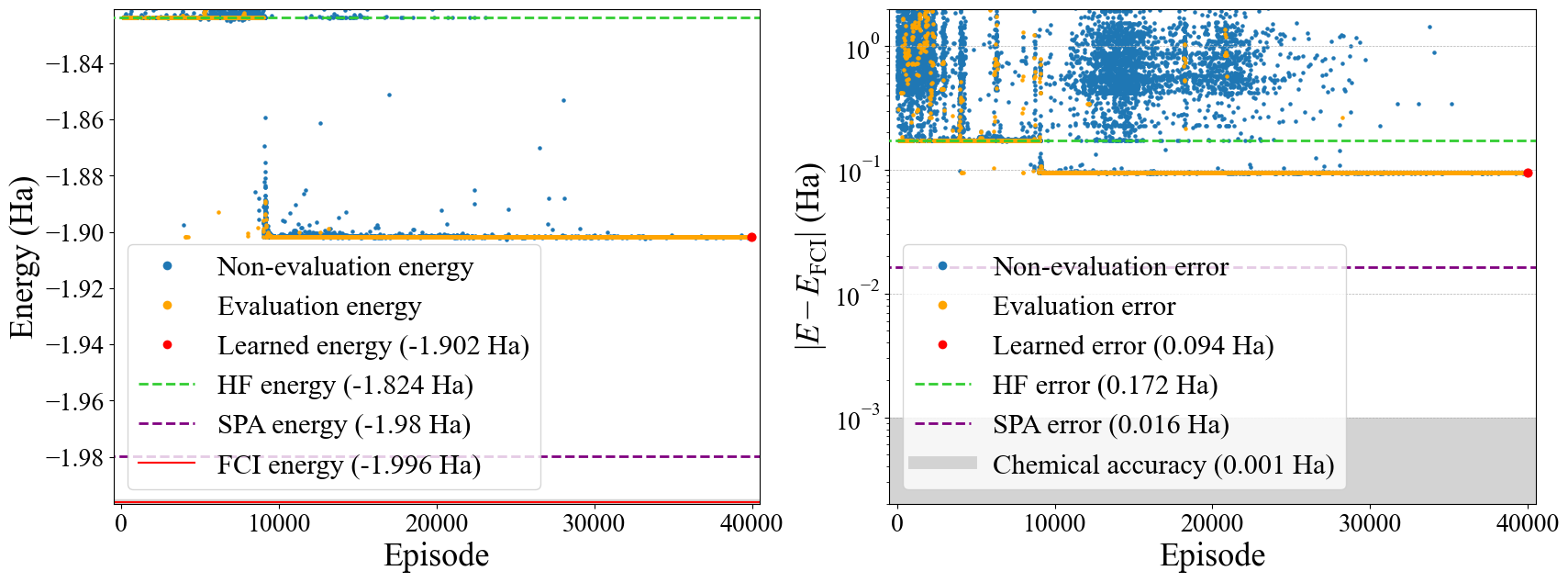}
    \caption{Left: Progression of the obtained energies during training for the best-case run of H$_4$ at a bond distance of 1.5\,\AA.
    Right: Energy error relative to the FCI energy during training, shown on a logarithmic scale, corresponding to the energies shown in the left panel.}
    \label{fig:best_case_H4}
\end{figure}

One best-case run is shown in Fig.~\ref{fig:best_case_H4}. The final learned energy for this run is $-1.9019$~Ha,  with an error of 0.0948~Ha relative to the FCI energy. 

\subsubsection{H\textsubscript{4} from 0.5\,\AA to 1.55\,\AA}
\label{app:H4_supplementary_material_PEC}
\begin{figure}[htbp]
    \centering
    \includegraphics[width=\textwidth]{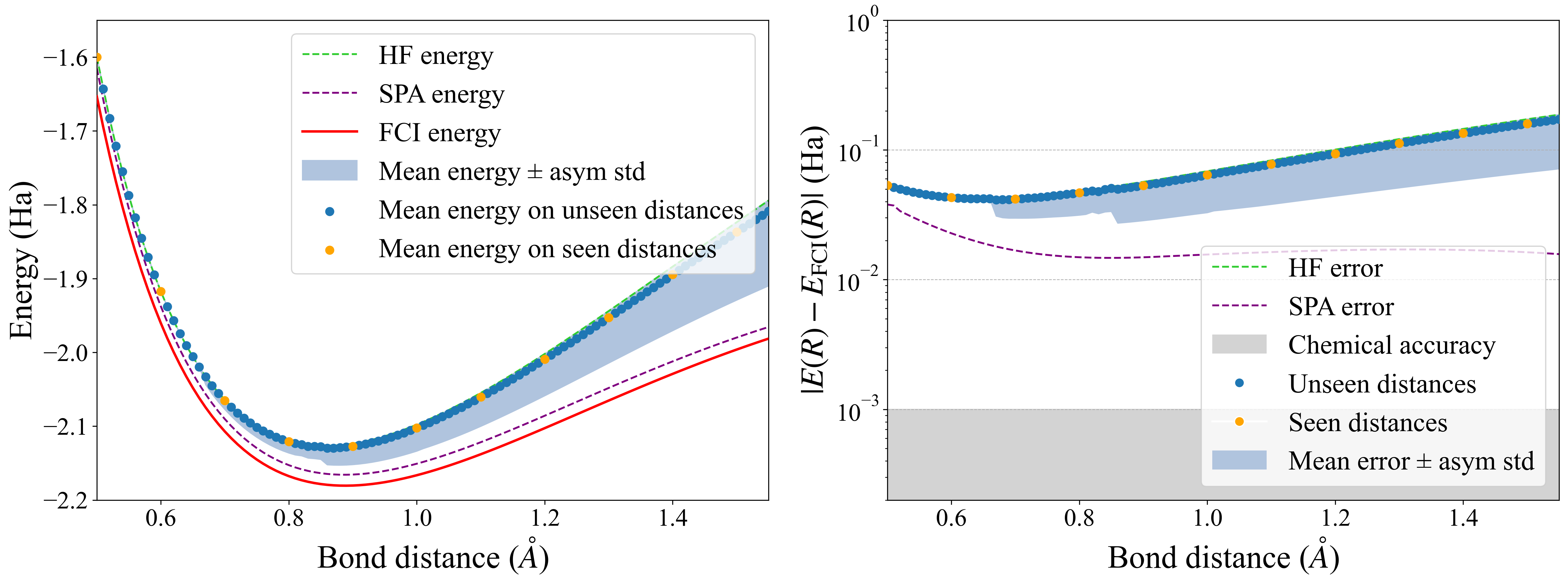}   
     \caption{Averaged PEC (left) and averaged error with respect to the FCI energy (right) across twelve runs for H$_4$. The shaded region represents the asymmetric standard deviation around the mean. The orange points represent the averaged learned energies and errors at the bond distances used during training. The blue points correspond to the averaged predicted energies and errors obtained for unseen bond distances, by generalization of the model.}
    \label{fig:PEC12_H4}
\end{figure}

For H$_4$ in the bond distance range 0.5--1.55\,\AA, we performed twelve independent runs using fixed hyper-parameters, listed in Appendix~\ref{app:hyper-parameters}. The hyper-parameter choices were guided by those previously shown to work well in the single bond distance setting (see Appendix~\ref{app:H4_supplementary_material_single}). The left panel of Fig.~\ref{fig:PEC12_H4} shows the mean PEC with one asymmetric standard deviation, while the right panel of Fig.~\ref{fig:PEC12_H4} displays the corresponding mean error relative to the FCI energy on a logarithmic scale. The orange energies and errors correspond to bond distances used during training, while the blue points indicate unseen bond distances, for which energy and error were predicted via generalization.
The average achieved error across the PEC is $0.0815^{+0.0056}_{-0.0381}$ Ha, with two runs outperforming HF across most bond distances.

\section{Framework details and implementation}
\label{app:implementation}

In the following, we provide additional details on the MDP framework introduced in Sec.~\ref{Theory} and describe the continuous-discrete Soft Actor-Critic (SAC) algorithm in detail.

\subsection{Details on MDP framework}

\subsubsection{Additional notes on state representation}
\label{app:state_representation}

In this framework, we adopt a state representation in which the quantum state is directly fed into the neural network. This contrasts with prior approaches that encode the gate-level description of the state itself as input to the neural network \cite{ostaszewskiReinforcementLearningOptimization2021}. Given the increased complexity of our setup, particularly the need to optimize continuous gate parameters and to generalize across varying Hamiltonians, we opted for a state-based encoding, as it introduces less ambiguity. In the gate-level description, different gate sequences can prepare the same quantum state, and the effect of a gate strongly depends on its position within the circuit—ambiguities that complicate learning and are alleviated mainly by using a representation of the quantum state as input.

\paragraph{Gaussian featurization.} The parameter $R\in[R_{\min},R_{\max}]$ is represented via a set of Gaussian basis functions, $g(R)=[g_1(R), g_2(R),\dots , g_J(R)]^T$ \cite{Hermann_2020} :
\begin{align}
    g_j(R)&=\exp\left[-\frac{1}{2}\left(\frac{R-\mu_j^G}{\sigma^G}\right)^2\right],\quad j=1,\dots ,J\label{gaussian_feat}\\
    \mu_j^G&=q_j,\quad \sigma^G=\frac{R_{\max}-R_{\min}}{J}, 
\end{align}
where $\{q_j\}_{j=1}^J$ are defined as $q_j=R_{\min}+\frac{j-1}{J-1}(R_{\max}-R_{\min})$, i.e., $J$ equally spaced points in the interval $[R_{\min},R_{\max}]$. 

\subsubsection{Additional notes on action encoding}\label{app:action_encoding}
The gate type and the qubits involved are encoded via an array-based action representation, as introduced by Ref.~\cite{Ostaszewski_2021}.
Each discrete action $d \in \{0, 1, \dots, |A|-1\}$, where $|A|$ denotes the total number of possible discrete actions, corresponds to an array with four entries encoding the gate and its target qubits:
\begin{align}
d: &\quad[\text{control qubit}, \text{target qubit}, \text{rotation qubit}, \text{rotation axis}]. \notag
\end{align}
The rotation axis is represented by integers, with $1 \rightarrow x$, $2 \rightarrow y$, and $3 \rightarrow z$. Qubits are indexed programmatically as $q \in \{0, \dots, n-1\}$ for an $n$-qubit system.
Special values are used to indicate the omission of certain gates, according to the following rules.
\begin{itemize}[topsep=6pt, itemsep=0pt, parsep=0pt]
    \item If the control qubit is set to $n$, no CNOT gate is applied, and the target qubit is set to $0$.
    \item If the rotation qubit is set to $n$, no single-qubit rotation is applied, and the rotation axis is set to $0$.
\end{itemize}
Combinations in which both the control and rotation qubits are set to $n$ (which would imply a no-op), or in which both are different from $n$ (which would imply a simultaneous CNOT and rotation), are not allowed in the discrete action set.

For example, for $n = 2$, which corresponds to $|A| = 8$ possible actions, the encoding takes the following form:
\vspace{-1em} 
\begin{center}
\setlength{\jot}{2pt}
\begin{alignat*}{3}
0: &\ [1,0,2,0]  &&\ \rightarrow\ \text{CNOT}(1,0) \\
1: &\ [0,1,2,0]  &&\ \rightarrow\ \text{CNOT}(0,1) \\
2: &\ [2,0,0,1]  &&\ \rightarrow\ \hat{R}_x(\theta)\ \text{on qubit 0} \\
3: &\ [2,0,0,2]  &&\ \rightarrow\ \hat{R}_y(\theta)\ \text{on qubit 0} \\
4: &\ [2,0,0,3]  &&\ \rightarrow\ \hat{R}_z(\theta)\ \text{on qubit 0} \\
5: &\ [2,0,1,1]  &&\ \rightarrow\ \hat{R}_x(\theta)\ \text{on qubit 1} \\
6: &\ [2,0,1,2]  &&\ \rightarrow\ \hat{R}_y(\theta)\ \text{on qubit 1} \\
7: &\ [2,0,1,3]  &&\ \rightarrow\ \hat{R}_z(\theta)\ \text{on qubit 1}
\end{alignat*}
\end{center}

Each discrete action is paired with a corresponding continuous action that specifies the gate parameter. The policy network, therefore, outputs a probability distribution over the discrete actions. For each discrete action, it specifies a mean $\mu_d$ and a standard deviation $\sigma_d$ to describe the associated continuous distribution for the parameters. Note that this, in principle, implies that CNOT is also associated with a continuous parameter; this is done solely to keep the network dimensions consistent and is ignored throughout the optimization procedure.

\subsubsection{Additional notes on reward design}\label{app:reward_design_details}

The reward function within our framework is defined as 
\begin{align} f(R;E_{t})&=c_{\exp}f_{\text{exp}}(R;E_{t})+c_{\text{lin}} f_{\text{lin}}(R;E_{t}),
\end{align}
which consists of an exponential and a linear component, weighted by pre-factors $c_{\exp}$ and $c_{\text{lin}}$ respectively.
The exponential component is defined as
\begin{align}
f(R; E_t) = \exp\left(-\frac{E_t - \mu_R}{\sigma_R}\right),
\end{align}
where the parameters $\mu_R$ and $\sigma_R$ are dynamically updated during training to adapt the center and slope of the exponential. They are computed from two energy buffers specific to each bond distance $R$:
\begin{itemize}
\item $\mathcal{E}_R^{\mu} = {E_{R,1}, \dots, E_{R,m}}$ contains the $m$ lowest energies observed for the value $R$ so far,
\item $\mathcal{E}_R^{\sigma} = {E_{R,m+1}, \dots, E_{R,m+k}}$ contains the next $k$ lowest energies for the value $R$.
\end{itemize}
Both sets are initialized with a single value $E_0$ and are progressively populated and updated after each training step. The parameters $\mu_R$ and $\sigma_R$ are then calculated as the average of these sets
\begin{align}
\mu_R &= \frac{1}{m} \sum_{i=1}^m E_i, \quad E_i \in \mathcal{E}_R^{\mu}, \label{mu_reward} \\
\sigma_R &= \left| \mu_R - \frac{1}{k} \sum_{j=1}^k E_j \right| + \sigma_{\min}, \quad E_j \in \mathcal{E}_R^{\sigma}, \label{sigma_reward}
\end{align}
where $\sigma_{\min}$ is a minimal bound to prevent the argument of the exponential function from diverging.
Note that, for simplicity, the main text presents the reward as if it were computed directly as an outcome of the environment.
In practice, however, the dynamic definition of the reward leads to a slight deviation from the standard MDP formulation, as transitions are stored in the replay buffer in the form $(s_t(R), a_t, E_t, E_{t+1}, s_{t+1}(R), \texttt{done}))$, where $E_t$ and $E_{t+1}$ denote the energies before and after applying action $a_t$ and $\texttt{done}$ is a binary flag to indicate whether $s_{t+1}(R)$ is a terminal state.
When a batch of transitions is sampled, the rewards are computed on-the-fly before the update step, based on the most recent values of $\mu_R$ and $\sigma_R$.
This dynamic adaptation ensures that energies are always evaluated relative to the RL agent's current best performance.
As energy estimates improve over time, the reward function's sensitivity to further improvements increases accordingly.
This is particularly important for capturing electronic correlations, which yield only marginal reductions in energy, yet are crucial for accurate ground-state characterization and therefore require strong reward signals to be reliably discovered.

The exponential reward component evaluates approximately to zero for energies $E_t(R) \gg \mu(R) + \sigma(R)$, resulting in a sparse reward signal in this regime. Such sparsity increases the likelihood that agents will get trapped in local minima~\cite{riedmiller2018learning}.
To mitigate this effect, a linear reward component $r_{\text{lin}}$ is added:
\begin{align}
r_{\text{lin}}(E_t, E_{t-1}) = -\left(E_t - E_{t-1}\right),
\end{align}
which independently provides an additional non-zero reward signal during training.

\subsection{Discrete-continuous soft actor-critic algorithm}
\label{SAC}

We use the \emph{soft actor-critic} (SAC) algorithm as the underlying RL framework to train our agent. The SAC algorithm is a model-free, off-policy actor-critic reinforcement learning algorithm introduced by Haarnoja et al.~\cite{haarnoja2018} and originally designed to learn a policy over continuous action spaces. For our purposes, which require handling both discrete and continuous actions $a=(d,c)$, we present a modified version as adapted by Ref.~\cite{Erdman_2023} to learn the joint probability distribution of the policy
\begin{align}
   \pi(c,d|s)=\pi_d(d|s)\cdot\pi_c(c|d,s),
\end{align}
where $\pi_d(d|s)$ describes the marginal probability of taking discrete action $d$, and $\pi_c(c|d,s)$ describes the conditional probability density of choosing action $c$, given action $d$.

The SAC algorithm incorporates an additional entropy-based reward term in its optimization objective, which has been shown to improve exploration efficiency and accelerate learning compared to state-of-the-art algorithms that rely solely on the standard RL objective \cite{haarnoja2018}:
\begin{align}
    \pi^*=\underset{\pi}{\argmax}\underset{}{\mathbbm{E}_{\pi}}\left[\sum_{k=1}^{T}\gamma^k\big(r_{k}+\alpha H(\pi(\,\cdot\,|s))\big)\,\bigg|\,s_0=\bar{s}\right].\label{sacobjective}
\end{align}
Since the entropy of the joint policy $\pi(c,d|s)$ decomposes as
$H[\pi(\cdot\mid s)] = H_d(s) + H_c(s)$, following \cite{Erdman_2023}, we replace the single-temperature term
$\alpha\, H[\pi(\cdot\mid s)]$ by $\alpha_d\, H_d(s) + \alpha_c\, H_c(s)$, with
\begin{align}
H_d(s) := H[\pi_d(\cdot\mid s)],\qquad
H_c(s) := \sum_{d} \pi_d(d\mid s)\, H\!\left[\pi_c(\cdot\mid d,s)\right],
\end{align}
where $\alpha=(\alpha_d,\alpha_c)\ge 0$ are “temperature” coefficients that control the exploration–exploitation balance (separately for the discrete and continuous parts).

As an actor-critic method, SAC learns both a policy (actor) and a value function (critic), with the critic evaluating the actor's actions to guide further optimization. 
The policy $\pi(c,d|s)$ is represented by a feed-forward neural network parametrized by the parameters $\phi=(\phi_d,\phi_c)$:
\begin{align}
   \pi_\phi(c,d|s)=\pi_{\phi_d}(d|s)\cdot\pi_{\phi_c}(c|d,s).
\end{align}
The continuous action distribution $\pi_{\phi_c}(c|d,s)$ employs a squashed Gaussian parameterization to enforce action bounds \cite{haarnoja2018,Erdman_2023}
\begin{align}
    \Tilde{u}_{\phi_c}(\varepsilon|d,s)=u_a+\frac{u_b-u_a}{2}[1+\tanh{(\mu_{\phi_c}(d,s)+\sigma_{\phi_c}(d,s)\cdot\varepsilon)}],\quad \varepsilon\sim\mathcal{N}(0,1),
\end{align}
where $\mu_{\phi_c}(d,s)$ and  $\sigma_{\phi_c}(d,s)$ represent the mean and standard deviation of the Gaussian distribution, and $\mathcal{N}(0,1)$ denotes the standard normal distribution with zero mean and unit variance. The lower and upper bounds $u_a$ and $u_b$ define the valid range of actions after applying the squashing function.

The value function $Q^\pi(s,a)$ estimates the value of taking action $a=(c,d)$ in state $s$ under the policy $\pi(c,d|s)$ and can be written as follows \cite{Erdman_2023} 
\begin{align}
    Q^\pi(s,a)=\mathbbm{E}_\pi\left[\sum_{k=1}^T\gamma^{k-1}\big(r_k+\alpha_d H_d^\pi(s_k)+\alpha_c H_c^\pi(s_k)\big)|  s_0=s,a_0=a\right].
\end{align}
The value function $Q^\pi(s,a)$ is represented by a feed-forward neural network parametrized by the parameters $\omega$. In practice, there are two value functions, $Q_{\omega_1}(s,a)$ and $Q_{\omega_2}(s,a)$, learned to counteract the overestimation bias of Q-values through the clipped double Q-learning trick~\cite{Fujimoto}. Additionally, there are two target value functions, $Q_{\omega_{\text{target},1}}(s,a)$ and $Q_{\omega_{\text{target},2}}(s,a)$, used to limit the impact of unfavorable parameter changes~\cite{Mnih}.

SAC starts from a random policy, runs the MDP with the current policy to collect transitions, stores them in the replay buffer, and then repeatedly alternates between two steps. The \emph{policy evaluation step} judges the performance of the current policy by learning the value function. The \emph{policy improvement step} utilizes the value function to enhance policy performance. As a third step, the temperature coefficients $(\alpha_d,\alpha_c)$ are automatically tuned to regulate exploration. This process is repeated to improve the policy step by step.  

\subsubsection{The policy evaluation step}
In the policy evaluation step, the parameters $\omega_1$ and $\omega_2$ of the value functions $Q_{\omega_1}(s,a)$ and $Q_{\omega_2}(s,a)$ are optimized towards minimizing the mean squared difference between the predicted return and an estimate for the actual return. This results in the following loss function for the Q-networks \cite{Erdman_2023} 
\begin{align} J_Q(\omega_i)=\mathbbm{E}_{(s,a,r,s'd)\sim\mathcal{D}}\left[\frac{1}{2}\big(Q_{\omega_i}(s,a)-y(r,s',d)\big)^2\right],\quad i\in\{1,2\},\label{JQ}
\end{align}
where $y(r,s',\texttt{done})$ is the so-called target,
\begin{align}
y(r,s',\texttt{done})=r+\gamma(1-\texttt{done})\underset{a'\sim\pi(\,\cdot\,|s')}{\mathbbm{E}}\left[\underset{j=1,2}{\min}Q_{\omega_{\text{target},j}}(s',a')+\alpha_d H_d(s')+\alpha_c H_c(s')\right].
\end{align}
The target $y(r,s',\texttt{done})$ represents the estimated actual return. It incorporates the immediate reward $r$, whose actual value is known from the batch's collected transitions. The expectation $\mathbbm{E}_{a'\sim\pi(\,\cdot\,|s')}[\cdot]$ in \eqref{JQ} can be calculated as \cite{Erdman_2023}
\begin{align}
    \mathbbm{E}_{a'\sim\pi(\,\cdot\,|s')}[\cdot]=\sum_{a'_d}\pi_d(a'_d|s')\underset{a'_c\sim\pi_c(\,\cdot\,|a'_d,s')}{\mathbbm{E}}[\cdot].
\end{align}
The parameters $\omega_{\text{target},1}$ and $\omega_{\text{target},2}$ of the target value functions $Q_{\omega_{\text{target},1}}(s,a)$ and $Q_{\omega_{\text{target},2}}(s,a)$ are updated via Polyak averaging~\cite{haarnoja2018}:
\begin{align}
   \omega_{\text{target,j}}\leftarrow \rho \omega_{\text{target,j}}+(1-\rho)\omega_{j},\quad j\in\{1,2\}, \label{polyak}
\end{align}
where the soft update factor $\rho\in[0,1]$ determines how fast the parameter changes of the value functions settle down to the target value functions.

\subsubsection{Policy improvement step}
In the policy improvement step, the policy parameters $\phi=(\phi_c, \phi_d)$ are trained to maximize the expected future return. For this purpose, the newly updated Q-networks from the policy evaluation step are used as guidance, as they provide improved estimates of the expected return and thereby indicate which actions are more likely to lead to higher rewards. The authors of Ref.~\cite{Erdman_2023} derive a loss function for the joint policy over discrete and continuous actions, which allows both components to be updated simultaneously using the following objective~\cite{Erdman_2023}:
\begin{align}
    J_\pi(\phi)=\mathbbm{E}_{s\sim\mathcal{B}}\left[\sum_d \pi_{\phi_d}(d|s)\left(\alpha_d\log\pi_{\phi_d}(d|s)+\alpha_c\log\pi_{\phi_c}(\tilde{u}_\phi(\varepsilon|d,s)|d,s)-\underset{j=1,2}{\min}Q_{\omega_j}(s,\tilde{u}_\phi(\varepsilon|d,s),d)\right)\right], \quad \varepsilon\sim\mathcal{N}(0,1).
\end{align}

\subsubsection{Temperature tuning}
The temperatures $\alpha_d$ and $\alpha_c$ are determined by minimizing two loss functions designed to regulate policy exploration throughout training according to Ref.~\cite{haarnoja2019} automatically:
\begin{align}
    J(\alpha_d)&=\alpha_d \underset{s\sim\mathcal{B}}{\mathbbm{E}}\big[H_d^{\pi}(s)-H^{\text{target}}_d\big]=\alpha_d\underset{s\sim\mathcal{B}}{\mathbbm{E}}\left[-\sum_d\pi_d(d|s)\log\pi_d(d|s)-H^{\text{target}}_d\right]
    , \label{lr_alpha_d}\\
    J(\alpha_c)&=\alpha_c \underset{s\sim\mathcal{B}}{\mathbbm{E}}\big[H_c^{\pi}(s)-H^{\text{target}}_c\big]=\alpha_c\underset{s\sim\mathcal{B}}{\mathbbm{E}}\left[-\sum_d\pi_d(d|s)\underset{c\sim\pi_c(\,\cdot\,|d,s)}{\mathbbm{E}}[\log\pi_c(c|d,s)]-H^{\text{target}}_c\right], \label{lr_alpha_c}
\end{align}
where $H_d^{\text{target}}$ and $H_c^{\text{target}}$ are the target entropies.
Similar to Ref.~\cite{Erdman_2023}, we chose the target entropies to follow an exponential decay over the course of training, defined as
\begin{align}
    H^{\text{target}}_{c,d}&=H^{\text{end}}_{c,d}+(H^{\text{start}}_{c,d}-H^{\text{end}}_{c,d})\exp\left(-\beta_{c,d}\frac{t_{\text{current}}}{t_{\text{total}}}\right),\label{def_target_e}\\
     t_{\text{current}}&=e_{\text{current}}\cdot T+t,\notag\\
    t_{\text{total}}&=e_{\text{max}}\cdot T,\notag
\end{align}
whereas $e_{\text{current}}$ denotes the current episode, $e_{\text{max}}$ the total number of training episodes, and $t$ the current step within the episode, with a maximum of $T$ steps.
The initial target entropies $H^{\text{start}}_{c,d}$ are set close to the maximal entropy to favor strong exploration in the initial training phase.
In contrast, the final target entropies $H^{\text{end}}_{c,d}$ are chosen to be comparatively small, thereby enforcing near-deterministic behavior. This setup ensures that the policy transitions from highly exploratory to increasingly deterministic behavior, with the rate of change governed by the decay factor $\beta_{c,d}$.

\subsubsection{Practical implementation details}
Updates of all network parameters are performed at regular intervals, defined by the update frequency $n_{\text{update}}$, during which $n_{\text{update}}$ consecutive updates are executed (as outlined above)~\cite{haarnoja2018,haarnoja2019}. 
For each update, a batch $\mathcal{B}$ of transitions of batch size $|\mathcal{B}|$ is uniformly sampled from the replay buffer $\mathcal{D}$. Sampling from the replay buffer enables approximating the expectation value in the SAC objective (Eq.~\eqref{sacobjective}), since the true state distribution is unknown and cannot be accessed directly.

\section{Application of the framework to optimization}
\label{app:optimization}
To showcase the versatile application domains of our framework, we demonstrate its application to the job-shop scheduling problem (JSP). For this purpose, we adopt the Hamiltonian job-shop scheduling formulation presented in \cite{Lucas_2014}, using the job lengths of a single job as a transferable parameter.
The corresponding Hamiltonian, as is typical in combinatorial optimization problems, is diagonal in the computational basis, implying that the ground state is a product state. Thus, the ground state switches discretely between orthogonal product states as the transferable job length changes. 
The discrete nature of ground state changes implies that transfer learning is inherently challenging, as neighboring solutions may lack structural similarity and cannot be smoothly adapted. Thus, while the presented setting may not initially appear to be a promising use case for our framework – particularly given its trivial nature due to the limited number of qubits – it nevertheless captures the essence of the optimization problem at a larger scale. In a broader setting, identifying a transferable approximate solution, e.g., as a warm start, could offer advantages in terms of computational cost and has a considerable practical interest. 

\paragraph{The problem formulation according to Ref.~\cite{Lucas_2014}.} Let $N$ be the number of jobs that must be executed in the shortest possible time on $m$ machines. Each job $i$ has an individual length $L_i \in \mathbbm{N}$, where all lengths are fixed except for the length of the last job, $L_N$, which denotes the transferable parameter in this setting. If a set of jobs $V_\alpha$ is assigned to the machine $\alpha$, the run time $M_\alpha$ of machine $\alpha$ is defined by $
    M_\alpha\coloneqq\sum_{i\in V_\alpha}L_i.$
Without loss of generality, machine 1 is defined as the longest running machine $M_1\geq M_\alpha\,\forall\,\alpha$. As the total run time $T_{\text{total}}$ is determined by the longest running machine, the task of the optimization is to minimize $\text{max}(M_1)$. 
This problem can be encoded within a Hamiltonian following Ref.~\cite{Lucas_2014} as
\begin{align}
    H_{\text{JSP}}(x,y)&=A\sum_{i=1}^N\big(1-\sum_{\alpha}x_{i,\alpha}\big)^2\notag +
        A\sum_{\alpha=1}^m\left(\sum_{n=1}^{\mathcal{M}}n y_{n,\alpha}+\sum_{i=1} L_i(x_{i,\alpha}-x_{i,1})\right)^2\notag+B\sum_{i=1}^N L_i x_{i,1}\label{JS}.
\end{align}
The first term enforces that each job $i$ is assigned to exactly one machine $\alpha$. The variable $x_{i,\alpha}$ indicates if job $i$ is added to machine $\alpha$
    \begin{align} 
        x_{i,\alpha}= \begin{cases}
                        1,\quad\text{if job $i$ is added to machine $\alpha$} ,\\
                        0,\quad\text{if job $i$ is not added to machine $\alpha$}.
                        \end{cases}
    \end{align} \\
The second term of $H_{\text{JSP}}$ encodes that the run time $M_1$ of machine $1$ should be the longest, i.e.,  $M_1\geq M_\alpha\,\forall \alpha$. The summation limit $\mathcal{M}$ is chosen by the user and describes the longest run-time difference between machine $\alpha$ to machine $1$. In the worst case, this can be $\mathcal{M}=\sum_i L_i$, corresponding to all jobs running on machine $1$. The user must trade off between a value that is too low, which may never satisfy the condition, and a value that is too high, which leads to easily satisfying the condition. The variable $y_{n,\alpha}$ indicates if the runtime difference $M_1-M_\alpha$ between machine 1 and machine $\alpha$ is $n$,
\begin{align}
    y_{n,\alpha}= \begin{cases}
                        1,\quad&\text{if}\quad M_1-M_\alpha=n,\\
                        0,\quad&\text{else}.
                        \end{cases}
\end{align}
The third summand encodes the minimization task of the total run time. The pre-factors $A$ and $B$ should should satisfy $0<B\cdot\text{max}(L_i)<A$.

To map the Hamiltonian \eqref{JS} to a qubit Hamiltonian, we require 
 $N\cdot m+ (m-1)\cdot\mathcal{M}$ qubits. In principle, it is also possible to reduce the number of needed qubits by using a logarithmic encoding for $\mathcal{M}$~\cite{Lucas_2014}. The qubits are assigned as
\begin{align}
\ket{q}
&= \ket{q_x} \otimes \ket{q_y} \notag\\
&= \Biggl(\;\bigotimes_{\alpha=1}^{m}\;\bigotimes_{i=1}^{N}\ket{q^{x}_{i,\alpha}}\Biggr)
   \;\otimes\;
   \Biggl(\;\bigotimes_{\alpha=2}^{m}\;\bigotimes_{n=1}^{\mathcal M}\ket{q^{y}_{n,\alpha}}\Biggr).
\end{align}

The $x_{i,\alpha}$ and $y_{n,\alpha}$ are binary variables which can be mapped to operators in the following way $x_{i,\alpha}\mapsto\, \frac{1}{2}(\hat{I}-\hat{\sigma}^z_{i,\alpha}) $, $y_{n,\alpha}\mapsto\, \frac{1}{2}(\hat{I}-\hat{\sigma}^z_{n,\alpha})$.

\paragraph{The demonstration setup.} For demonstration purposes, we consider the following setup: Let $N=3$ jobs be assigned to $m=2$ machines. Jobs $j=1,2$ have a fixed processing time of $L_{1,2}=1$, while the time of job $j=3$ has a range of $L_3\in\{1,2,3\}$.
The maximum allowed runtime difference between the two machines is set to $\mathcal{M}=1$. The pre-factors of the Hamiltonian are chosen as $B=1$ and $A=4$. This setup results in a seven-qubit Hamiltonian. The constructed circuits were restricted to a maximum of twelve gates, and the initial state
vector has been chosen as $\ket{0}^{\otimes n}$. Similar to the chemistry case, we executed twelve runs using the hyper-parameters specified in Table~\ref{tab:hyperparams}.

\begin{figure}[t!]
        \centering
       \includegraphics[width=0.9\textwidth]{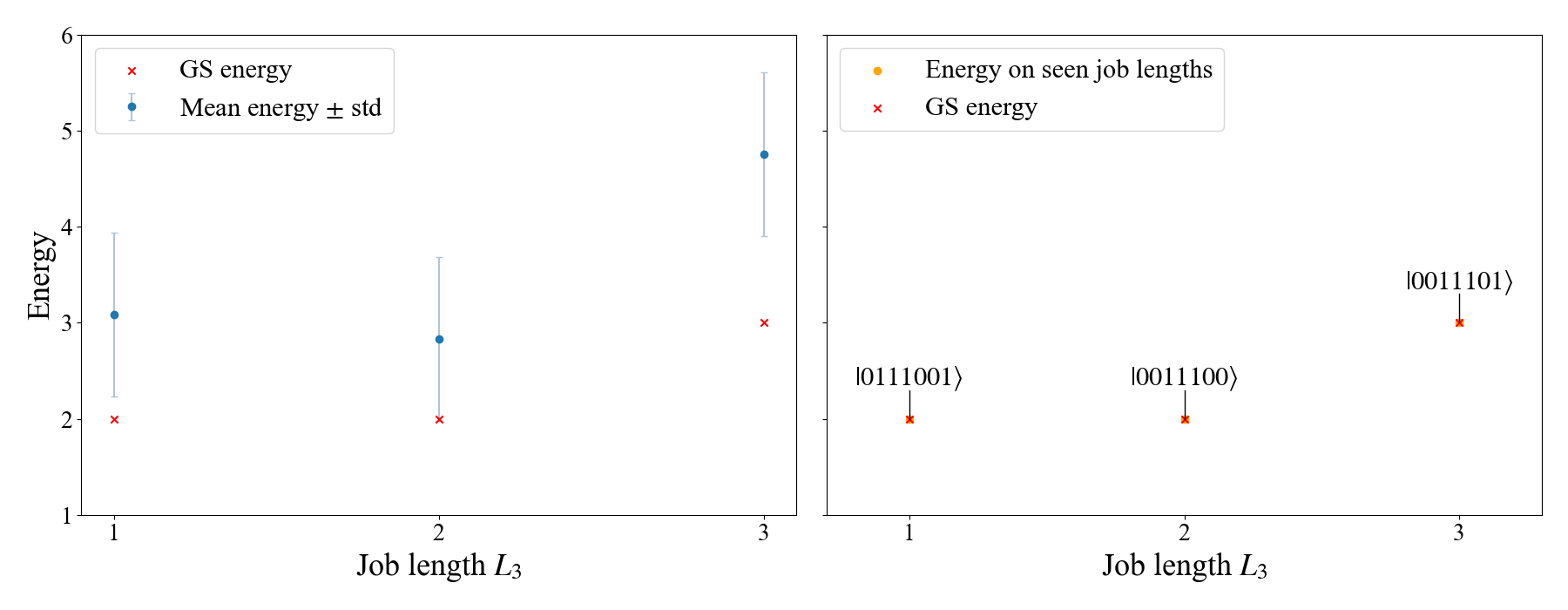}
    \caption{Left: Average performance for the JSP setup across the twelve runs. Right: Best-case performance for the JSP setup, including the learned states for each job length. }
    \label{fig:jsp_performance}
\end{figure}

\paragraph{Results.} The left of Figure~\ref{fig:jsp_performance} shows the average achieved energy $\pm$ one standard deviation across all three trained job lengths. Out of the twelve runs, four successfully recovered the exact solution for all three job lengths. The right panel presents one representative best-case solution and also includes the three learned states. These were slightly approximated, as the obtained states represent a superposition of the true ground state and a negligible contribution from an additional state. 

\begin{figure}[t!]
        \centering
        \includegraphics[width=0.25\textwidth]{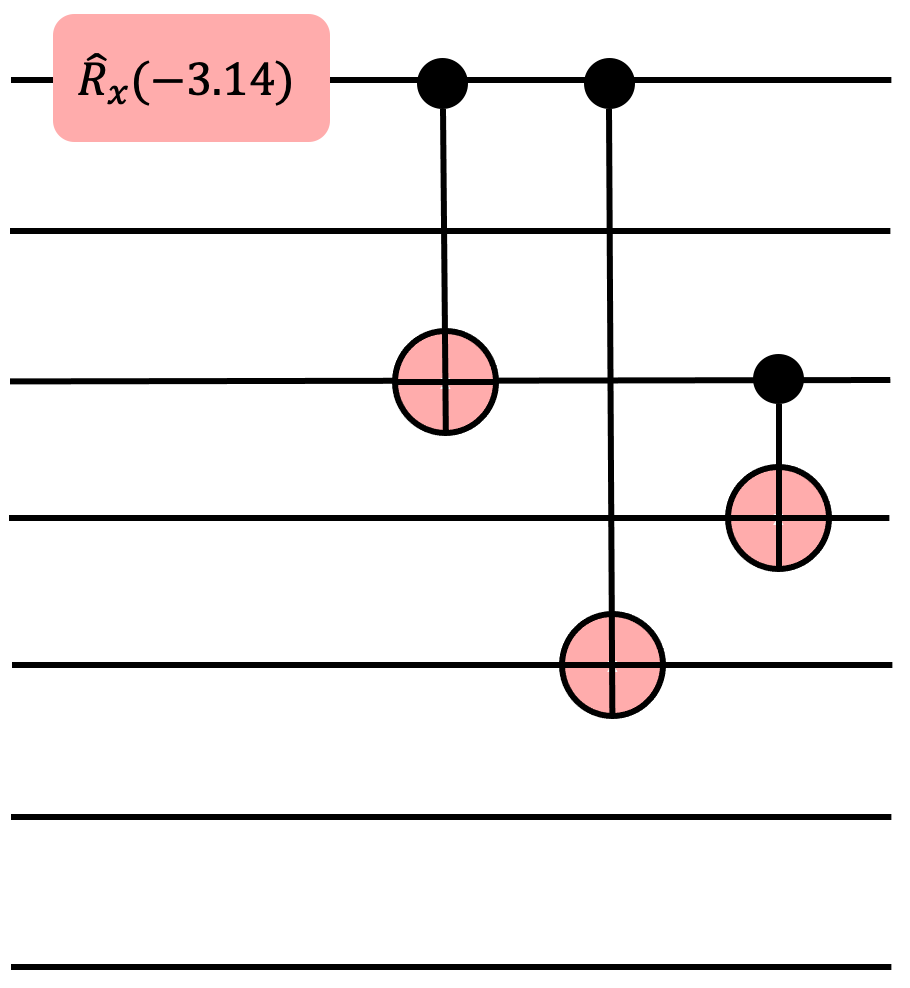}
    \caption{Example circuit for $L_3=3$ after further optimization as described in Appendix~\ref{circuits_pre-processing}. The circuit achieves the exact energy of $E=3$.}
    \label{fig:jsp_circuit}
\end{figure}

Additionally, in Figure~\ref{fig:jsp_circuit} we show an exemplary circuit discovered by the agent, after applying the pre-processing described in Appendix~\ref{circuits_pre-processing}, for the job length $L_3$.
To conclude, we demonstrated that our framework also applies to settings beyond quantum chemistry calculations. While the considered example is limited in scale, if extended to higher qubit numbers, the agent might be capable of identifying transferable solution patterns that generalize across varying job lengths and could serve as efficient warm starts for more exact optimization methods. However, we emphasize that the scalability of this approach to industry-relevant problem sizes remains an open question.

\section{Specification of molecule systems} \label{app:quantumchem}

Table~\ref{molecule_configurations} provides an overview of the molecular systems and the specific settings used to generate the qubit Hamiltonians. A notable technical subtlety when generalizing the quantum circuit across bond distances is the occurrence of orbital phase flips at seemingly arbitrary bond lengths, caused by the freedom in the SCF initial-guess orbitals \cite{jensen2007introduction}. While the ground-state energy remains invariant and the orbitals remain physically equivalent up to a global phase, this induces discontinuities in the wavefunction along the PEC. Since a smoothly varying ground state is desirable for evaluating the framework performance, we explicitly monitored the evolution of the Hamiltonian coefficients as a function of bond distance and manually corrected orbital phases where necessary to ensure continuity. In our case, this issue occurred for the four- and six-qubit LiH at $2.5$\,\AA, which was manually corrected.

\begin{table}[ht]
\footnotesize
\centering
\begin{tabular}{|>{\columncolor{blue!10}\raggedright\arraybackslash}p{4.8cm}!{\vrule width 1pt}>{\centering\arraybackslash}p{3.2cm}|>{\centering\arraybackslash}p{3.2cm}!{\vrule width 1pt}>{\centering\arraybackslash}p{5.0cm}|}
\hline
\multicolumn{1}{|c!{\vrule width 1pt}}{} & \textbf{Four-qubit LiH} & \textbf{Six-qubit LiH} & \textbf{H$_4$} \\
\specialrule{1.5pt}{0pt}{0pt} 
Qubit number  & 4 & 6 & 8 \\
\hline
Geometry  &Li: (0, 0, 0);  H: (0, 0, $R$) &Li: (0, 0, 0); H: (0, 0, $R$) & H: (0, 0, 0), (0, 0, $R$), (0, 0, $2R$), (0, 0, $3R$) \\
\hline
Bond distance $R$ &  $R \in [1, 4]$ & $R \in [1, 4]$ &  $R \in [0.50, 1.55]$ \\
\hline
Basis & STO-3G & STO-3G & STO-3G  \\
\hline
Active orbitals & $Li-2s$, $Li-2pz$, $H-1s$& $Li-2s$, $Li-2pz$, $H-1s$& full active space \\
\hline
Mapping & Parity mapping & Jordan–Wigner mapping & Jordan–Wigner mapping \\
\hline
Additional orbital optimization & no & no & yes\textsuperscript{1} \\
\hline
HF state & $\ket{1100}$\textsuperscript{2} & $\ket{100100}$\textsuperscript{2} & $\ket{11001100}$\textsuperscript{2} \\
\hline
Used chemistry package & Tequila\textsuperscript{3} and Qiskit Nature\textsuperscript{4}& Tequila\textsuperscript{3} and Qiskit Nature\textsuperscript{4} & Tequila\textsuperscript{3} \\
\hline
\specialrule{1.5pt}{0pt}{0pt}
\end{tabular}
\caption{Overview of the system configurations for each molecule. 
\textsuperscript{1}The orbital optimization for H$_4$ is applied exactly as described in Ref.~\cite{Kottmann_2023}.
\textsuperscript{2}Hartree-Fock (HF) states follow the blocked spin-ordering convention, with qubits numbered from $0$ to $n$. 
\textsuperscript{3}See \cite{Kottmann_2021, tequila_github_2023} for Tequila.
\textsuperscript{4}See \cite{qiskit_nature_2023} for Qiskit Nature. }
\label{molecule_configurations}
\end{table}

\section{Hyper-parameters}\label{app:hyper-parameters}

Table~\ref{tab:hyperparams} lists all parameters used in the numerical simulations presented in this work.

\begin{table}[ht]
\footnotesize
\centering
\begin{tabular}{|>{\columncolor{blue!10}\raggedright\arraybackslash}p{4.8cm}!{\vrule width 1pt}>{\centering\arraybackslash}p{1.6cm}|>{\columncolor{gray!10}\centering\arraybackslash}p{1.6cm}!{\vrule width 1pt}>
{\centering\arraybackslash}p{1.6cm}|>{\columncolor{gray!10}\centering\arraybackslash}p{1.6cm}!{\vrule width 1pt}>{\centering\arraybackslash}p{1.6cm}|>{\columncolor{gray!10}\centering\arraybackslash}p{1.6cm}!{\vrule width 1pt}>{\centering\arraybackslash}p{1.6cm}|}
 \hline
\multicolumn{1}{|c|}{\textbf{Hyper-parameters}} & 
\multicolumn{2}{|c|}{\textbf{Four-qubit LiH}} & 
\multicolumn{2}{|c|}{\textbf{Six-qubit LiH}} & 
\multicolumn{2}{|c|}{\textbf{H$_4$}} &
\textbf{JSP} \\
\specialrule{1.5pt}{0pt}{0pt} 
 \hline
\multicolumn{8}{|c|}{System settings} \\
\specialrule{1.5pt}{0pt}{0pt} 
 Bond distance interval & 2.2\,\AA & [1.0,4.0]\,\AA & 2.2\,\AA &[1.0,4.0]\,\AA & 1.5\,\AA&[0.5,1.6]\,\AA & \{1,2,3\} \\
\hline
 Bond distance step size training & - & 0.1 &-&0.1&-&0.1 & 1 \\
 \hline
  Bond distance step size prediction & - & 0.01  &-&0.01&-&0.01 & -\\
\hline
 Initial energy $E_0$& -7.0 Ha&  -7.0 Ha & -7.0 Ha &-7.0 Ha& 0.0 Ha& 0.0 Ha & 30\\
 \hline
 Initial state $\psi_0$& $\ket{1100}$ &  $\ket{1100}$& $\ket{001001}$& $\ket{001001}$&$\ket{11001100}$& $\ket{11001100}$ & $\ket{0000000}$\\
\specialrule{1.5pt}{0pt}{0pt} 
\multicolumn{8}{|c|}{Training settings} \\
\specialrule{1.5pt}{0pt}{0pt} 
 Maximal number of episodes $e_{\text{max}}$  & 30,000 & 40,000&40,000& 50,000&40,000& 50,000 & 45,000 \\
 \hline
 Maximal number of gates $T$ & 12 & 12 & 12&12 &10& 10 &12 \\
  \hline
Evaluation:non evaluation ratio & 10:1 &10:1& 10:1& 10:1& 10:1& 10:110:1 & \\
  \hline
Runs & 12 & 12& 12& 12 & 12 & 12 & 12\\
\specialrule{1.5pt}{0pt}{0pt} 
\multicolumn{8}{|c|}{SAC parameter settings} \\
\specialrule{1.5pt}{0pt}{0pt} 
 Learning rate critic $\lambda_q$ & 0.001 &0.001&0.001&0.001&0.001&0.001 &0.001 \\
 \hline
 Learning rate actor $\lambda_\pi$ &   0.001  &0.001&0.001&0.001&0.001&0.001 & 0.001\\
 \hline
 Learning rate alpha discrete $\lambda_{\alpha_d}$ & 0.003 &0.003&0.003&0.003&0.003&0.003 & 0.003\\
 \hline
 Learning rate alpha continuous $\lambda_{\alpha_c}$ & 0.003 &0.003&0.003&0.003&0.003&0.003 & 0.003\\
 \hline
 Batch size $|B|$ & 512 &512&512&512&512&512 &512 \\
 \hline
 Replay buffer capacity $|\mathcal{D}|$ &  18,000 &36,000&24,000& 36,000&50,000& 100,000 & 24,00 \\
 \hline
 Soft update factor $\rho$ & 0.005 &0.005& 0.005&0.005 &0.005&0.005 & 0.005 \\
 \hline
 Training update factor $n_{\text{update}}$ & 50 &50&50 &50&50&50 & 50\\
 \hline
 Initial random steps $n_{\text{random}}$ & 1,200 &1,200& 1,200 &1,200&1,000&1,000 &1,200 \\
 \hline
 Gamma $\gamma$ & 1 &1& 1 &1&1&1 &1 \\
 \hline
\specialrule{1.5pt}{0pt}{0pt} 
\multicolumn{8}{|c|}{Target entropies settings} \\
\specialrule{1.5pt}{0pt}{0pt} 
 Decay factor discrete target entropy $\beta_d$ & 1.2& 1.0&1.25 & 1.0 &1.0&1.0 & 1.0\\
 \hline
 Decay factor continuous target entropy $\beta_c$ & 2.1 & 2.0& 2.25&2.0&2.0&2.0 & 2.0\\
 \hline
 Deduction discrete target entropy  & 0.1& 0.1& 0.3& 0.4&0.4&0.4 & 0.1\\
 \hline
 Deduction continuous target entropy  & 0.05 &0.05&0.1&0.2&0.2&0.2 & 0.05\\
 \hline
 Target end value discrete entropy $H^{\text{end}}_d$ & 0.5& 0.5&0.5 & 0.5&1&1 & 0.5\\
 \hline
 Target end value continuous entropy  $H^{\text{end}}_c$  & -2 &-2&-2&-2&-1&-1 &-2 \\
\specialrule{1.5pt}{0pt}{0pt} 
\multicolumn{8}{|c|}{Neural network settings} \\
\specialrule{1.5pt}{0pt}{0pt}  
 Hidden layers actor &   256, 128 &  256, 128 & 256, 128& 256, 128& 256, 128& 256, 128 & 256, 128 \\
 \hline
  Hidden layers critic &   256, 128, 128 & 256, 128, 128 &256, 128, 128 &256, 128, 128 &256, 128, 128&256, 128, 128 & 256, 128, 128\\
\specialrule{1.5pt}{0pt}{0pt} 
\multicolumn{8}{|c|}{Reward settings} \\
\specialrule{1.5pt}{0pt}{0pt}  
 $\sigma_{\min}$  & 0.01 & 0.01 &0.005&0.0125&0.01&0.0075 &0.25 \\
 \hline
 $m$  & 15 & 30  &50&30&50&30 &30 \\
 \hline
 $k$  & 30 &50&70&50&70&50 & 50\\
 \hline
pre-factor exponential reward $c_{\exp}$ & 5 & 5 &5& 5&5&5 & 5\\
 \hline
pre-factor linear reward  $c_{\text{lin}}$ &  0.1 & 1 & 0.1 &0.1&0.1&0.1 & 1\\
\specialrule{1.5pt}{0pt}{0pt} 
\multicolumn{8}{|c|}{Gaussian featurization settings} \\
\specialrule{1.5pt}{0pt}{0pt} 
Number of embeddings $J$& -& 3 & -&6 &-&6 & 3\\
\hline
$[a,b]$& - & [1,4]& -&[1,4]&-&[0.5,2] &[0.9,3.1] \\
\hline
\end{tabular}
\caption{Overview of all parameter settings within the framework for the numerical simulations of all setups.}
\label{tab:hyperparams}
\end{table}

\end{document}